\documentclass[12pt,a4paper]{article}

\usepackage[english]{babel}
\usepackage{amsmath,amsthm,amssymb}
\usepackage{bbm,graphicx,psfrag,cite}

\numberwithin{equation}{section}

\newcommand{\cte}{\mathrm{Const}}
\newcommand{\Id}{\mathbbm{1}}
\newcommand{\Or}{\mathcal{O}}
\newcommand{\Z}{\mathbbm{Z}}
\newcommand{\R}{\mathbbm{R}}
\newcommand{\Pb}{\mathbbm{P}}
\newcommand{\Hilb}{\mathcal{H}}
\newcommand{\dx}{\mathbf{d}}
\newcommand{\sgn}{\mathrm{sgn}}
\renewcommand{\Re}{\mathrm{Re}}

\newcommand{\Ai}{\mathrm{Ai}}
\newcommand{\Af}{{\cal A}_{\rm 1}}
\newcommand{\e}{\varepsilon}
\newcommand{\Dtn}[3]{\frac{\dx^{#3} #1}{\dx #2^{#3}}}
\newcommand{\Dt}[2]{\frac{\dx #1}{\dx #2}}
\newcommand{\I}{{\rm i}}

\newtheorem{prop}{Proposition}[section]
\newtheorem{thm}[prop]{Theorem}
\newtheorem{lem}[prop]{Lemma}
\newtheorem{defin}[prop]{Definition}
\newtheorem{cor}[prop]{Corollary}
\newtheorem{rem}[prop]{Remark}

\newenvironment{proofOF}[2]{\removelastskip\vspace{6pt}\noindent
 {\it Proof of #1.}~\rm#2}{\qed \par\vspace{6pt}}

\numberwithin{equation}{section}

\title{Fluctuations in the discrete TASEP with periodic initial configurations and the Airy$_1$ process}
\author{Alexei Borodin\thanks{California Institute of Technology, e-mail: borodin@caltech.edu},
Patrik L. Ferrari\thanks{Technische Universit\"at M\"unchen, e-mail: ferrari@ma.tum.de},
Michael Pr\"ahofer\thanks{Technische Universit\"at M\"unchen, e-mail: praehofer@ma.tum.de}}

\date{25th November 2006}

\begin{document}
\maketitle \sloppy

\begin{abstract}
We consider the totally asymmetric simple exclusion process (TASEP) in discrete time with sequential update. The
joint distribution of the positions of selected particles is expressed as a Fredholm determinant with a kernel defining a signed determinantal point process. We focus on periodic initial conditions where particles occupy $d\Z$, $d\geq 2$. In the proper large time scaling limit, the fluctuations of particle positions are described by the Airy$_1$ process. Interpreted as a growth model, this confirms universality of fluctuations with flat initial conditions for a discrete set of slopes.
\end{abstract}

\section{Introduction}
The totally asymmetric simple exclusion process (TASEP) in discrete time consists of particles on $\Z$ with at most one particle at each site (exclusion principle). At each time step, particles jump to the neighboring right site with probability $p\in (0,1)$, provided the target site is empty. There are mainly two types of update rules one can consider. One is called \emph{parallel update}. It consists in first checking for all particles if they can jump (i.e., if their right neighbor is empty) and then, simultaneously and independently, these particles jump to the right each with probability $p$. The second update rule, the one we actually analyze in this paper, is called \emph{sequential update}. In this case, at each time step particles are processed sequentially from right to left, i.e., starting from right to left, if the site on the right of a particle is empty, then this particle jumps there with probability $p$. This update rule allows to shift blocks of particles to the right during one time-step. Equivalently, it can be regarded as a parallel update rule for holes, which under the constraint of keeping their order, jump to the left $k$ steps with probability proportional to $p^k$.

On a macroscopic level the particle density, $u(x,t)$, evolves deterministically according to the Burgers equation $\partial_t u + \partial_x(u(1-u)/(p^{-1}-u))=0$. It is therefore natural to focus on fluctuations, which turn out to have unexpected features. For example, the fluctuations of particle positions live on a $t^{1/3}$ scale and the limiting distribution depends on the type of the initial configuration.

In this paper we consider the positions of particles as observables. If one starts with step initial
conditions, i.e., particles occupying $\Z_-$, then the macroscopic particle density has a region of linear
decay~\cite{R81}. There the fluctuations are governed by the GUE Tracy-Widom distribution~\cite{TW94}, see~\cite{Jo00b}. Moreover, in the limit of large time $t$, the joint distributions of particle positions are governed by the Airy process~\cite{PS02} (called Airy$_2$ process in the following). The second situation which has been analyzed is the stationary initial condition~\cite{FS05a}. These results have been obtained for the continuous time and/or parallel update discrete time TASEP, but can be easily extended to the sequencial update case.

A third class of initial conditions consists in deterministic and periodic initial configurations. In~\cite{Sas05} and~\cite{BFPS06} the continuous time TASEP has been studied for the particular case of alternating initial configuration, i.e., initially with particles on $2\Z$. In this case the one-particle fluctuations are described by the GOE Tracy-Widom distribution. Joint distributions are given by a Fredholm determinant and the kernel converges pointwise to the one of the Airy$_1$ process. In the present paper we extend these results in several directions. Firstly, we look at the fully discretized version of the TASEP with sequencial update. Secondly, our result holds for a wider set of periodic initial configurations and, thirdly, the control on the kernel is stronger and we prove convergence of joint distribution functions. The corresponding results for continuous time can be obtained along the same lines.

The Airy$_1$ process, $\Af$, is a marginal of the signed determinantal point process with extended kernel $K_{\rm F_1}$, in the same way as the Airy$_2$ process is related to the extended Airy kernel. Explicitly, we set $B_0(x,y)=\Ai(x+y)$ and let $\Delta$ denote the one-dimensional Laplacian. Then
\begin{equation}
K_{\rm F_1}(u_1,s_1;u_2,s_2)=-(e^{(u_2-u_1)\Delta})(s_1,s_2)\Id(u_2>u_1)+(e^{-u_1\Delta} B_0 e^{u_2\Delta})(s_1,s_2).
\end{equation}
More explicitly, as shown in Appendix A of~\cite{BFPS06}, one has
\begin{eqnarray}\label{eqKernelExpanded}
& &\hspace{-2em} K_{\rm F_1}(u_1,s_1;u_2,s_2)=-\frac{1}{\sqrt{4\pi (u_2-u_1)}}\exp\left(-\frac{(s_2-s_1)^2}{4 (u_2-u_1)}\right) \Id(u_2>u_1) \nonumber \\
& &\hspace{-2em} + \Ai(s_1+s_2+(u_2-u_1)^2) \exp\left((u_2-u_1)(s_1+s_2)+\frac23(u_2-u_1)^3\right).
\end{eqnarray}
In particular, the one-point distribution of the {Airy$_1$ process} is given in terms of the GOE Tracy-Widom distribution~\cite{TW96}, as shown in~\cite{FS05b},
\begin{equation}
\Pb(\Af(0)\leq s)=F_1(2s).
\end{equation}

The process $\Af$ was first described by Sasamoto~\cite{Sas05}. The starting point is a determinantal formula for the probability distribution of the continuous time TASEP with finitely many particles, first discovered by Sch\"utz~\cite{Sch97}. Sasamoto found a clever reformulation making possible the large time asymptotic analysis for the positions of particles. The details of the derivation are presented in~\cite{BFPS06}. There we show how the signed determinantal point process arises by applying a result on the L-ensembles defined in~\cite{RB04}.

For the discrete-time sequential update TASEP, a determinantal formula of the same type as in~\cite{Sch97} was derived in~\cite{RS05} via the Bethe Ansatz. Since this is the starting point of our analysis we present here an elementary derivation. Very recently, a determinantal formula for parallel update was obtained~\cite{PP06}. Whether or not in this case a similar approach can be used for asymptotic analysis has still to be investigated.

By universality, one expects that the same limit process appears
for any deterministic and periodic initial configurations, not
only for particles initially on $2\Z$. In this paper we consider
initial conditions with one particle every $d$ sites. The
fluctuations of the position of a particle live on the scale
$t^{1/3}$, while the positions of two particles are non-trivially
correlated (on the $t^{1/3}$ scale) if they are of order $t^{2/3}$
apart. In Theorem~\ref{ThmConvToSasProc} we prove that, properly
rescaled, the joint distributions of particle positions converge
to the joint distributions of the Airy$_1$ process. That is, let
$x_k(t)$ be the position at time $t$ of the particles starting at
$-d k$, then for some constants $\mu,\sigma,\xi_1,\xi_2$ depending
only on $p$ and $d$, one has
\begin{equation}
\frac{x_{[\xi_1 t+\xi_2 u t^{2/3}]}(t)-\mu u t^{2/3}}{\sigma t^{1/3}}\longrightarrow\Af(u),
\quad\textrm{ as }t\to\infty,
\end{equation}
in the sense of finite dimensional distributions.

\begin{figure}[t!]
\begin{center}
\psfrag{x}[c]{$x$}
\psfrag{h}[l]{$h_t(x)$}
\includegraphics[height=3cm]{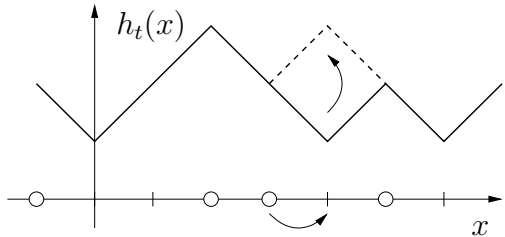}
\caption{The growth model associated to the TASEP. If a particle jumps to the right, then the surface growth vertically as indicated by the dashed line.}\label{FigGrowthModel}
\end{center}
\end{figure}

\vspace{1em} \emph{Reformulation of the result}. The TASEP can be interpreted as a stochastic growth model of an interface, which turns out to belong to the Kardar-Parisi-Zhang (KPZ) universality class introduced in~\cite{KPZ86}. Given a realization of the particle process, one defines for each time $t$ a continuous height function, $x\mapsto h_t(x)$, as follows. The height at $0$ is given by $h_t(0)=2 N_t$, where $N_t$ is the number of particles which jumped from site $0$ to site $1$ during the time interval $[0,t)$. If at time $t$ a particle is at site $x$, then $h_t$ decreases linearly with slope $-1$ in the interval $[x,x+1]$, and increases with slope $+1$ if the site is empty. Thus, if $x$ is an integer, one has always $|h_t(x+1)-h_t(x)|=1$. A particle jumping at time $t$ from $x$ to $x+1$ corresponds to increasing $h_t(x+1)$ by $2$. The surface obtained in this way grows vertically under the TASEP dynamics as indicated in Figure~\ref{FigGrowthModel}.

The step-initial conditions, corresponding to $h_0(x)=|x|$,  generate a curved macroscopic shape and the limit process is the Airy$_2$ process, first described in another growth model, the polynuclear growth model~\cite{PS02}. The alternating initial conditions, particles starting from $2\Z$, correspond to growth on a flat and horizontal substrate, $h_0(x)=(1+(-1)^x)/2$, $x\in\Z$. In this case the macroscopic limit shape is a constant function and the fluctuations are governed by the Airy$_1$ process. By universality one expects that the result are unchanged if the initial substrate is flat but \emph{tilted}. Theorem~\ref{ThmConvToSasProc} confirms universality for a discrete set of non-zero slopes (corresponds to $d\geq 3$).

The TASEP can also be interpreted as a directed percolation model, where ``flat'' corresponds to the point-to-line, while ``curved'' to the point-to-point setting, see e.g.~\cite{Jo00b,FerPhD,Pra03}. Finally, it has been conjectured~\cite{Fer04,BFPS06} that the evolution of the largest eigenvalue of GOE Dyson's Brownian Motion of random matrices is governed by the Airy$_1$ process too. This conjecture is however not based on KPZ universality.

\vspace{1em} \emph{New features}.
In the analysis some new interesting features appear. The kernel can be expressed in a double integral representation and, as stated in Theorem~\ref{ThmKernel}, one of the contour integrals circumscribes $d-1$ simple poles, which are the roots of a polynomial. There is only one simple case, $d=2$, which is equivalent to flat and horizontal initial conditions. In this case, there is only one pole and the asymptotic analysis reduces to the one of a single integral on the complex plane. An explicit solution of the polynomial equation does not exist for arbitrary $d\geq 3$, thus we had to employ several new ideas to circumvent the problem. Similar situations might appear in other problems like point-to-line directed percolation and all sorts of growth models on a flat substrate.

\vspace{1em} \emph{Outline}. The remainder of the paper is organized as follows. In Section~\ref{SectModelResult} we define the model precisely and state the main result of this work. In Section~\ref{SectDetMeasure} we give a new derivation of the determinantal formula of~\cite{RS05} and then apply our previous work~\cite{BFPS06} to the discrete-time TASEP. In Section~\ref{SectOrtho} we perform the necessary orthogonalization and obtain the finite time kernel. The asymptotic analysis is the content of Section~\ref{SectAsympt}. In Appendix~\ref{AppTraceClass} we deal with the trace-class problem of the kernel $K_{\rm F_1}$. In Appendix~\ref{SectAppendix} we describe an alternative, more constructive, way of obtaining the orthogonalizazion.

\subsection*{Acknowledgment}
A. Borodin was partially supported by the NSF grant DMS-0402047 and the CRDF grant RIM1-2622-ST-04.
P.L. Ferrari thanks M. Loss for sketching the proof that the one-point kernel $B_0$ is trace-class during his visit to Munich in 2005.

\section{Model and results}\label{SectModelResult}
The model analyzed in this paper is the discrete-time TASEP with sequential update on $\Z$. At any given time $t$, every site $j\in \Z$ can be occupied at most by one particle. Thus a configuration of the TASEP can be described by
$\eta=\{\eta_j,j\in\Z|\eta_j\in\{0,1\}\}$. $\eta_j$ is called the \emph{occupation variable} of site $j$, which is defined by $\eta_j=1$ if site $j$ is occupied and $\eta_j=0$ if site $j$ is empty.

\begin{figure}[t!]
\begin{center}
\psfrag{p1}[c]{$1-p$}
\psfrag{p2}[c]{$p(1-p)$}
\psfrag{p3}[c]{$p^2$}
\includegraphics[width=0.8\textwidth]{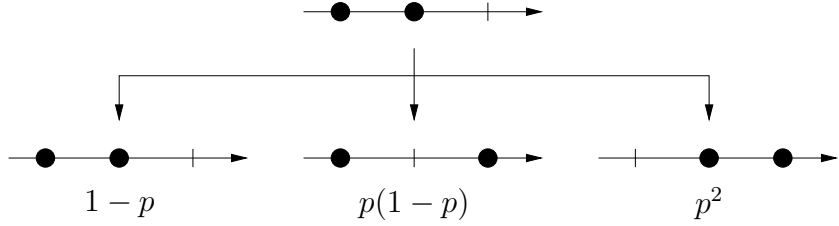}
\caption{Example of the transition weights with sequential update. Above is the configuration at time $t$. Below are the possible outcomes at time $t+1$ together with the transition weights.}\label{FigSequentialDynamics}
\end{center}
\end{figure}

Let $\eta(t)$ be a TASEP configuration at time $t$. Then the configuration at time $t+1$ is obtained by the following dynamics. Starting from right to left, a particle jumps to the neighboring site with probability $p\in (0,1)$ provided this site is empty. Since the update is sequential from right to left, during a time step a block of consecutive particles can jump. For example, if at time $t$ we have $\eta_0(t)=\eta_1(t)=1$ and $\eta_2(t)=0$, then the particle at $0$ can jump to $1$ provided the particle at $1$ jumps to $2$, see Figure~\ref{FigSequentialDynamics}.

Let us start at time $t=0$ with particles occupying the sublattice
$d\Z$. The observable we focus at is the joint distribution of
particle positions, which can be expressed as a Fredholm determinant
as stated in Theorem~\ref{ThmKernel}.
\begin{thm}\label{ThmKernel}
Let $\sigma(1)<\sigma(2)<\ldots<\sigma(m)$ be the indices of the $m$ particles starting at time $t=0$ from positions $x_k(0)=-d(\sigma(k)-1)$. Denote by $x_k(t)$ their positions at time $t$. Their joint distribution is given by
\begin{equation}
\Pb\Big(\bigcap_{k=1}^m \big\{x_{\sigma(k)}(t) \geq a_k\big\}\Big)=
\det(\Id-\chi_a K\chi_a)_{\ell^2(\{\sigma(1),\ldots,\sigma(m)\}\times\Z)},
\end{equation}
where $\chi_a(\sigma(k),x)=\Id(x<a_k)$ and the kernel $K$ is defined as follows.

For fixed $v$, let $u_0(v),\ldots,u_{d-1}(v)$ be the roots of the equation $u(1+u)^{d-1}=v(1+v)^{d-1}$, with $u_0(v)=v$ the trivial solution. Then the kernel for the $d$-spaced initial configuration is
\begin{equation}\label{eq285}
K(n_1,x_1;n_2,x_2)= -\binom{x_1-x_2-1}{n_2-n_1-1} + K_0(n_1,x_1;n_2,x_2)
\end{equation}
with
\begin{eqnarray}\label{eq285b}
& &K_0(n_1,x_1;n_2,x_2)\\
&=&\frac{1}{2\pi \I}\oint_{\Gamma_0}\dx v\sum_{i=1}^{d-1} \frac{1+dv}{1+d u_i(v)} \frac{(1+pu_i(v))^t (-u_i(v))^{n_1} (1+v)^{x_2+n_2-2}}{(1+pv)^t (-v)^{n_2} (1+u_i(v))^{x_1+n_1+1}}\nonumber,
\end{eqnarray}
where $\Gamma_0$ is any simple loop anticlockwise oriented around the pole at $v=0$ (without other poles being inside $\Gamma_0$).
\end{thm}
For alternating initial configurations, $d=2$, the kernel is particularly simple.
\begin{cor}\label{CorollaryTwoPeriodic}
The kernel for the $2$-periodic configuration is given by
\begin{eqnarray}\label{eqKernelD2}
K(n_1,x_1;n_2,x_2)&=&-\binom{x_1-x_2-1}{n_2-n_1-1}\\
& & +\frac{-1}{2\pi \I}\oint_{\Gamma_0} \dx v \frac{(1+v)^{x_2+n_1+n_2-2}}{(-v)^{x_1+n_1+n_2-1}}\left(\frac{1-p-pv}{1+pv}\right)^t. \nonumber
\end{eqnarray}
\end{cor}
\begin{proof}
Apply (\ref{eq285b}) with $u_1(v)=-1-v$.
\end{proof}

In Theorem~\ref{ThmConvToSasProc} we prove the convergence of the
fluctuations of particle positions to the {Airy$_1$ process}. Thus
we need to define this process and the scaling limit.
\begin{defin}[The Airy$_1$ process]
Let $B_0(x,y)=\Ai(x+y)$, $\Delta$ the one-dimensional Laplacian, and the kernel $K_{\rm F_1}$ defined by
\begin{equation}\label{eqKernelF1}
K_{\rm F_1}(u_1,s_1;u_2,s_2)=-(e^{(u_2-u_1)\Delta})(s_1,s_2)\Id(u_2>u_1)+(e^{-u_1\Delta} B_0 e^{u_2\Delta})(s_1,s_2).
\end{equation}
The \emph{{Airy$_1$ process}} $\Af$ is the process with $m$-point joint distributions at \mbox{$u_1< u_2< \ldots <u_m$} given by the Fredholm determinant
\begin{equation}\label{eq2.4}
\Pb\Big(\bigcap_{k=1}^m\{\Af(u_k)\leq s_k\}\Big)=
\det(\Id-\chi_s K_{\rm F_1}\chi_s)_{L^2(\{u_1,\ldots,u_m\}\times\R)},
\end{equation}
where $\chi_s(u_k,x)=\Id(x>s_k)$.
\end{defin}

\begin{rem}{\rm
The kernel $K_{\rm F_1}$ is not trace-class on $L^2(\{u_1,\ldots,u_m\}\times\R)$, because the diffusion part appearing for $u_2>u_1$ makes it not even Hilbert-Schmidt. However, as shown in Appendix~\ref{AppTraceClass}, there exists a conjugate operator, one with a kernel leading to the same Fredholm expansion of \mbox{$\det(\Id-\chi_s K_{\rm F_1}\chi_s)$}, that is trace-class on $L^2(\{u_1,\ldots,u_m\}\times\R)$. Thus the Fredholm determinant in (\ref{eq2.4}) regarded as its Fredholm expansion series is well defined.}
\end{rem}

We focus on the region around the origin. The fluctuations of the particle positions are of order $t^{1/3}$ and non-trivial correlations occur for particles at a distance of order $t^{2/3}$, as expected from the KPZ scaling exponents. The mean density of particles is $1/d$ and the probability of jumping to the neighboring site is $p$, provided the position is available. With sequential update, a particle can jump to the right even if the target site is occupied, provided that the blocking particle itself jumps during the same time step. Therefore in the stationary state the probability that the right position is available is given by
\begin{equation}
\sum_{k\geq 0} (p/d)^k \frac{d-1}{d}=\frac{d-1}{d-p}.
\end{equation}
The average speed of particles is $p(d-1)/(d-p)$, which means that at time $t$ particles with index close to $t p(d-1)/((d-p)d)$ will be close to the origin. This motivates the following scaling.\\

\noindent {\bf Scaling limit}. Define the constants
\begin{equation}\label{eqS1}
\kappa=\frac{(2(1-p)p)^{1/3}(d(d-1))^{2/3}}{d-p},\quad \mu=-\kappa^2 \frac{2}{d-1}.
\end{equation}
We consider particles with index $n(u,t)$ given by
\begin{equation}\label{eqS2}
n(u,t)=\bigg[\frac{p(d-1)}{d(d-p)}t-\frac{\mu}{d} u t^{2/3}\bigg]
\end{equation}
which typically at time $t$ is at a position close to $\mu u t^{2/3}$. Recall that the initial condition is $x_k(0)=-d(k-1)$, $k\in\Z$.
Then the rescaled process is given by
\begin{equation}\label{eqRescaled}
u\mapsto X_t(u)=\frac{x_{n(u,t)}-\mu u t^{2/3}}{-\kappa t^{1/3}}
\end{equation}
and converges to the Airy$_1$ process as follows.

\begin{thm}\label{ThmConvToSasProc}
Let $X_t$ be the rescaled process as in (\ref{eqRescaled}). Then, in the limit of large time $t$, it converges to the Airy$_1$ process $\Af$,
\begin{equation}
\lim_{t\to\infty} X_t = \Af,
\end{equation}
in the sense of finite-dimensional distributions (given by (\ref{eq2.4})).
\end{thm}

\section{Signed determinantal point process}\label{SectDetMeasure}
We first consider a system with a finite number $N$ of particles. The \mbox{$d$-periodic} configuration is then obtained by the proper $N\to\infty$ limit. We start at time $t=0$ with $N$ particles at positions $y_N < \ldots < y_2 < y_1$ and study the probability that at time $t$ these particles are at positions $x_N < \ldots < x_2 < x_1$. Denote this transition probability by
\begin{equation}
G_t(x_1,\ldots,x_N)=\Pb((x_N,\ldots,x_1;t)|(y_N,\ldots,y_1;0)).
\end{equation}

\begin{lem}[See \cite{RS05}]\label{lem1}
The transition probability has a determinantal form
\begin{equation}\label{eqGreen}
G_t(\{x\})=\det(F_{i-j}(x_{N+1-i}-y_{N+1-j},t))_{1\leq i,j \leq N}
\end{equation}
with
\begin{equation}\label{eqFn}
F_{n}(x,t)=(1-p)^t \frac{(-1)^n}{2\pi \I} \oint_{\Gamma_{0,1}} \frac{\dx w}{w} \left(1+\frac{p}{1-p} w\right)^t \frac{(1-w)^{-n}}{w^{x-n}},
\end{equation}
where $\Gamma_{0,1}$ is any simple loop around $0$ and $1$ oriented anticlockwise.
\end{lem}
\textbf{Notational remark:} with the notation $\int_{\Gamma_S}f(z)\dx z$ with $S$ a set of points we mean that the integral is taken over any simple loop, oriented anticlockwise, enclosing the set $S$, and such that \emph{all} the poles of $f$ inside $\Gamma_S$ belong to $S$. Equivalently, $\int_{\Gamma_S}f(z)\dx z=2\pi \I \sum_{z\in S}\mathrm{Res}(f,z)$.

This result is already contained in~\cite{RS05}, where the derivation used the Bethe-ansatz method. Here we present a different derivation.
\begin{proofOF}{Lemma~\ref{lem1}}
To obtain (\ref{eqGreen}) we first write the master equation for the $N$ particles,
\begin{equation}\label{eqMaster}
G_{t+1}(x_N,\ldots,x_1)=\hspace{-1em}\sum_{b_1,\ldots,b_N\in\{0,1\}} (1-p)^N \left(\frac{p}{1-p}\right)^{b_1+\ldots+b_N} \hspace{-1em} G_{t}(x_N-b_N,\ldots,x_1-b_1)
\end{equation}
with the boundary conditions due to the exclusion constraint
\begin{equation}\label{eqBC}
G_t(x_N,\ldots,x_{k+1},x_{k+1}+1,x_{k-1},\dots,x_1)\!=\!G_t(x_N,\ldots,x_{k+1},x_{k+1},x_{k-1},\ldots,x_1),
\end{equation}
for $k=1,\ldots,N-1$, cp.\ with~\cite{Sch97}.

We assume that a formula of the form (\ref{eqGreen}) holds and prove that there exists a family of functions $\{F_n\}$ satisfying the Lemma. As an abbreviation we set $F_{n}^{j}(x,t)=F_{n}(x_{N+1-j}-y_{N+1-j},t)$ during the proof. Inserting (\ref{eqGreen}) into the master equation (\ref{eqMaster}) and using the multi-linearity of the determinant, we get
\begin{eqnarray}
& & G_{t+1}(x_N,\ldots,x_1)\nonumber \\
&=&\hspace{-1em}  \sum_{b_1,\ldots,b_N\in\{0,1\}} (1-p)^N \left(\frac{p}{1-p}\right)^{b_1+\cdots+b_N}\hspace{-1em} \det\left[F_{i-j}^{j}(x_{N+1-i}-b_{N+1-i},t)\right]_{1\leq i,j \leq N}\nonumber \\
&=& \det\left[(1-p)\sum_{b_{N+1-i}=0}^1\left(\frac{p}{1-p}\right)^{b_{N+1-i}} \hspace{-1em} F_{i-j}^{j}(x_{N+1-i}-b_{N+1-i},t)\right]_{1\leq i,j \leq N}\\
&=&\det\left[(1-p) F_{i-j}^{j}(x_{N+1-i},t)+p F_{i-j}^{j}(x_{N+1-i}-1,t)\right]_{1\leq i,j \leq N}.\nonumber
\end{eqnarray}
This is equal to $\det\left[F_{i-j}^{j}(x_{N+1-i},t+1)\right]_{1\leq i,j \leq N}$ if the functions $F_n$'s satisfy
\begin{equation}\label{eqRel1}
F_n(x,t+1)=(1-p) F_n(x,t)+p F_n(x-1,t).
\end{equation}

We also have to take into account the boundary conditions. The two sides of (\ref{eqBC}) are $N\times N$ determinants with all the rows identical except for the $(N+1-k)$th one. Combining the two determinants, (\ref{eqBC}) reads
\begin{equation}
0=\det\left[\begin{array}{c} \vdots \\ F_{N-k-j}^{j}(x_{k+1},t)\\ F_{N+1-k-j}^{j}(x_{k+1},t)-F_{N+1-k-j}^{j}(x_{k+1}+1,t)\\ \vdots \end{array}\right]_{1 \leq j \leq N}.
\end{equation}
The two lines explicitly written are linearly dependent whenever the functions $F_n$ satisfy
\begin{equation}\label{eqRel2}
F_{n-1}(x,t)=c(F_n(x,t)-F_n(x+1,t))
\end{equation}
for some $c$. Here we choose $c=1$. Any other choice of $c$ corresponds to the replacement $F_n$ by $c^{-n} F_n$, which, however, keep the determinant in (\ref{eqGreen}) unchanged.

The functions $F_n$ are determined by the two relations (\ref{eqRel1}) and (\ref{eqRel2}), as well as the initial condition
\begin{equation}\label{eq3.10}
G_0(x_N,\ldots,x_1)=\delta_{y_N,x_N}\cdots \delta_{y_1,x_1}.
\end{equation}

$F_0(x,t)$ is already determined by one-particle configurations. In fact, in this case, $G_t(x)=\Pb(x(t)=x|x(0)=y)=F_0(x-y,t)$. Therefore
\begin{equation}
F_0(x-y,t)=(1-p)^t\left(\frac{p}{1-p}\right)^{x-y} \binom{t}{x-y}.
\end{equation}
This result is consistent with (\ref{eqRel1}) and (\ref{eq3.10}). Denote by $\Delta$ the discrete derivative $\Delta f(x)\equiv f(x+1)-f(x)$. Then (\ref{eqRel2}) implies
\begin{equation}
F_{-n}(x,t)=(-1)^n(\Delta^n F_0)(x,t).
\end{equation}
$F_0$ has the following integral representation,
\begin{equation}
F_0(x,t)=(1-p)^t \frac{1}{2\pi \I} \oint_{\Gamma_0} \frac{\dx w}{w} \left(1+\frac{p}{1-p} w\right)^t \frac{1}{w^x},
\end{equation}
where $\Gamma_0$ is any simple loop around $0$ oriented anticlockwise. Therefore to obtain $F_{-n}$ we simply apply
\begin{equation}
\Delta^n \frac{1}{w^x}=\frac{(1-w)^n}{w^{n+x}}.
\end{equation}
Thus, for $n\geq 0$,
\begin{equation}\label{eq3.15}
F_{-n}(x,t)=(1-p)^t \frac{(-1)^n}{2\pi \I} \oint_{\Gamma_0} \frac{\dx w}{w} \left(1+\frac{p}{1-p} w\right)^t \frac{(1-w)^{n}}{w^{x+n}}.
\end{equation}
In this case, there is no pole at $w=1$, thus replacing $\Gamma_0$ by $\Gamma_{0,1}$ leaves the result unchanged.

For $n>0$, $F_n$ is determined by the recurrence relation
\begin{equation}\label{RecRel}
F_{n+1}(x,t)=\sum_{y\geq x} F_n(y,t)
\end{equation}
together with the property that $F_0(x,t)=0$ for $x$ large enough. To have the sum in
(\ref{RecRel}) well defined we need $|w|>1$, i.e., the integration
path includes $1$, which is a pole of the integrand for $n\geq 1$. Thus in order
for (\ref{RecRel}) to be satisfied for all $n$, we need to take
into account the poles both at $0$ and $1$.
\end{proofOF}

The analogue of Theorem~\ref{ThmKernel} for a finite number of particles is the following Proposition, which is a consequence of Lemma 3.4 of~\cite{BFPS06}.
\begin{prop}\label{propKernel}
Let $\sigma(1)<\sigma(2)<\ldots<\sigma(m)$ be the indices of $m$ out of the $N$ particles initially at $y_{\sigma(1)},\ldots,y_{\sigma(m)}$.
The joint distributions of their positions $x_{\sigma(k)}(t)$ is given by
\begin{equation}
\Pb\Big(\bigcap_{k=1}^m\left\{x_{\sigma(k)}(t)\leq a_k\right\}\Big)=
\det(\Id-\chi_a K\chi_a)_{\ell^2(\{\sigma(1),\ldots,\sigma(m)\}\times\Z)}
\end{equation}
where $\chi_a(\sigma(k),x)=\Id(x>a_k)$. $K$ is the extended kernel with entries
\begin{equation}\label{eqKernelFinal}
K(n_1,x_1;n_2,x_2)=-\phi^{(n_1,n_2)}(x_1,x_2)+\sum_{i=0}^{n_2-1} \Psi^{n_1}_{n_1-n_2+i}(x_1) \Phi^{n_2}_{i}(x_2)
\end{equation}
where
\begin{equation}
\phi^{(n_1,n_2)}(x_1,x_2) = \binom{x_1-x_2-1}{n_2-n_1-1}.
\end{equation}
The functions $\Psi_i^n$, $n\geq 1$, $i\in\Z$, are defined by
\begin{equation}\label{eqPsi}
\Psi_i^{n}(x)=(1-p)^t \frac{1}{2\pi \I} \oint_{\Gamma_{0}} \frac{\dx w}{w^{n+1}} \left(1+\frac{p}{1-p} w\right)^t \frac{(1-w)^{n}}{w^{x-y_{n-i}}}
\end{equation}
where the path $\Gamma_0$ in the definition of $\Psi_i^n$ is any simple loop, anticlockwise oriented, which includes the pole at $w=0$ but not the one at $w=1$. The functions $\Phi_i^n$, $i=0,\ldots,n-1$, $n\geq 1$, are polynomials of degree at most $n-1$ uniquely defined by
\begin{equation}\label{eqOrtho}
\sum_{x\in\Z}\Phi_i^{n}(x) \Psi_j^{n}(x)=\delta_{i,j},\quad j=0,\ldots,n-1.
\end{equation}
\end{prop}
\begin{proof} The proof is an application of Lemma 3.4 of~\cite{BFPS06}, with
\begin{equation}
\phi_{n}(x_i^{n},x_j^{n+1})= \Id(x_i^{n}>x_j^{n+1}),\quad n=1,\ldots,N-1,
\end{equation}
and
\begin{equation}
\Psi^{N}_{N-i}(x)=(-1)^{N-i} F_{-N+i}(x-y_{i},t),\quad i=1,\ldots,N.
\end{equation}
The above functions $F_i$ are defined by an integral enclosing $w=0$ only, since $w=1$ is not a pole for $i\leq 0$.
With the definition (\ref{eqPsi}) above we have the composition rule
\begin{equation}\label{eq3.25a}
(\phi*\Psi^{n+1}_{n+1-j})(x)=\Psi^n_{n-j}(x).
\end{equation}
In our setting, if we sum up all the variables $\{x_j^m,1\leq m <n,1\leq j \leq m\}$, we obtain a Vandermonde determinant in the variables $x_j^n$. Thus the space $V_n$ of Lemma 3.4 in~\cite{BFPS06} is generated by $\{1,x,\ldots,x^{n-1}\}$ and the $\Phi^n_k$ are polynomials of order at most $n-1$. A simple computation using (\ref{eqPsi}) leads to
\begin{equation}
\sum_x \Psi_{j}^{n}(x) = \begin{cases}0, & j=1,\ldots,n-1, \\ 1, & j=0,\end{cases}
\end{equation}
which, together with (\ref{eqOrtho}) leads to $\Phi^n_0(x)=1=\phi_{n-1}(\infty,x)$. Then the Proposition~\ref{propKernel} follows from Lemma 3.4 of~\cite{BFPS06}.
\end{proof}

\section{Orthogonalization}\label{SectOrtho}
Consider the case where the particles are initially regularly spaced as follows
\begin{equation}
y_i=-d(i-1),\quad i=1,\ldots,N.
\end{equation}
for $d\geq 2$. For any fixed time $t$, we first obtain the kernel for a fixed number of particles, $N$. Then we take the $N\to\infty$ limit and focus on a point far enough to the left where the particles do not feel the fact that there is a rightmost particle.

\begin{lem}\label{lemOrtho}
The functions $\Psi^{n}_k(x)$ and $\Phi^{n}_k(x)$ have the following integral representations. Let $z=x+d(n-1)$, then
\begin{equation}\label{eqFctPsi}
\Psi^n_k(x)=\frac{(-1)^k}{2\pi \I}\oint_{\Gamma_0}\frac{\dx w}{w^{z+1}} (1+p(w-1))^t ((w-1)w^{d-1})^k
\end{equation}
and
\begin{equation}\label{eqFctPhi}
\Phi^n_k(x)= \frac{(-1)^k}{2\pi \I}\oint_{\Gamma_0}\frac{\dx v}{v} \frac{1+dv}{(1+pv)^t}\frac{(1+v)^{z-1}}{(v(1+v)^{d-1})^{k}}
\end{equation}
where $\Gamma_0$ is any anticlockwise simple loop enclosing only the pole at $0$.
\end{lem}
\begin{proof}
We have
\begin{equation}
\Psi^n_k(x)=(-1)^k F_{-k}(x-y_{n-k},t)= (-1)^k F_{-k}(z-dk,t).
\end{equation}
Thus (\ref{eqFn}) leads directly to (\ref{eqFctPsi}).
Next we prove that (\ref{eqFctPhi}) satisfies the orthogonality relation (\ref{eqOrtho}). Since $\Psi^n_k(x)=0$ for $x<-d(n-1)$, i.e.\ $z<0$, we have
\begin{eqnarray}\label{eq246}
& &\sum_{z\geq 0}\Psi^n_k(x(z))\Phi^n_j(x(z)) =\frac{(-1)^k}{2\pi \I} \oint_{\Gamma_0}\dx w (1+p(w-1))^t \big((w-1)w^{d-1}\big)^k\nonumber \\
& & \times \frac{(-1)^j}{2\pi \I}\oint_{\Gamma_0} \frac{\dx v}{v} \frac{(1+dv)}{(1+pv)^t (v(v+1)^{d-1})^j} \sum_{z\geq 0}\frac{(v+1)^{z-1}}{w^{z+1}}
\end{eqnarray}
provided that the integration domain satisfies $|1+v| < |w|$.
The last sum gives
\begin{equation}
 \sum_{z\geq 0}\frac{(v+1)^{z-1}}{w^{z+1}}=\frac{1}{(w-(1+v))(1+v)}.
\end{equation}
(\ref{eq246}) has a simple pole at $w=1+v$. Therefore the integration over $w$ leads to
\begin{equation}\label{eq248}
\sum_{z\geq 0}\Psi^n_k(x(z))\Phi^n_j(x(z))= \frac{(-1)^{k+j}}{2\pi \I} \oint_{\Gamma_0}\dx v \frac{1+dv}{v(1+v)} (v(1+v)^{d-1})^{k-j}.
\end{equation}
The final step is a change of variable. Let $u=v(1+v)^{d-1}$. Then
\begin{equation}
\dx u = (1+v)^{d-2}(1+dv)\dx v
\end{equation}
and the integral is again around $0$. Thus
\begin{equation}
\sum_{z\geq 0}\Psi^n_k(x(z))\Phi^n_j(x(z))= \frac{(-1)^{k+j}}{2\pi \I} \oint_{\Gamma_0}\dx u \frac{1}{u^{j+1-k}}=\delta_{j,k}.
\end{equation}
\end{proof}

There is also a more constructive way of doing the orthogonalization using Krawtchouk orthogonal polynomials. In fact, as shown in Appendix~\ref{SectAppendix}, the $\Psi^{(N)}$'s can be written as linear combinations of Krawtchouk polynomials on the interval $[0,t+d(N-1)]$ and parameter $p$. What one has to do is to invert a certain $N\times N$ matrix with entries depending on the initial conditions. We were able to do it in the case $d=2$ and obtain the formula of Lemma~\ref{lemOrtho}. Once the form of the $\Phi$'s is obtained for $d=2$, it is easy to find the proper ansatz in the case $d>2$.

\begin{proofOF}{Theorem~\ref{ThmKernel}}
We apply Proposition~\ref{propKernel} with the orthogonalization of Lemma~\ref{lemOrtho}. With the change of variable $u=w-1$ in (\ref{eqFctPsi}) the function $\Psi^n_k$ has an integral representation with a pole at $u=-1$, namely
\begin{equation}\label{eq4.10}
\Psi^n_k(x)=\frac{(-1)^k}{2\pi \I}\oint_{\Gamma_{-1}}\frac{\dx u}{(1+u)^{z+1}}(1+pu)^t(u(1+u)^{d-1})^k
\end{equation}
with $z=x+d(n-1)$. The main part of the kernel, denoted by $K_0(n_1,x_1;n_2,x_2)$, reads
\begin{eqnarray}\label{eq42}
& & \sum_{k=0}^{n_2-1}\Psi^{n_1}_{n_1-n_2+k}(x_1)\Phi^{n_2}_k(x_2)=\frac{(-1)^{n_1-n_2}}{(2\pi \I)^2}\oint_{\Gamma_0}\dx v \frac{(1+dv)(1+v)^{z_2+d-2}}{(1+pv)^t(v(1+v)^{d-1})^{n_2}}\nonumber \\
& & \times \oint_{\Gamma_{-1}} \dx u
\frac{(1+pu)^t (u(1+u)^{d-1})^{n_1}}{(1+u)^{z_1+1}} \frac{1}{u(1+u)^{d-1}-v(1+v)^{d-1}}
\end{eqnarray}
with \mbox{$z_i=x_i+d(n_i-1)$} provided that the paths of integration for $u$ and $v$ satisfy (a) $|u(1+u)^{d-1}|>|v(1+v)^{d-1}|$ and that (b) $u=0$ is not inside the contour $\Gamma_{-1}$. The expression (\ref{eq42}) is obtained as follows from (\ref{eqKernelFinal}). First we exchange the finite sum and the integrals. Then we extend the sum over $k$ to $-\infty$. The expression remains unchanged if the two above conditions are satisfied. In fact, for $k\leq -1$, the pole at $v=0$ vanishes because of the extra term $v^{-k}$ and by (a) the series is absolutely summable. Condition (b) reflects the fact that in the definition of (\ref{eqPsi}) the integration path includes only the pole at $0$. A possible choice of the integration paths is $\Gamma_{-1}=\{u, |1+u|=1/2\}$ and $\Gamma_0$ a contour with $|v|$ small enough to satisfy (a).

The kernel for the $d$-spaced initial configuration is obtained from the finite system by choosing the $x_i$ such that the kernel becomes independent of the fact that we have only a finite number of particles. In other words, we choose $x_i$ such that the kernel is invariant with respect to the shift of the positions by $-d$ and the particle numbers by $1$. This is achieved when $z_i<(d-1)n_i$ because $u=-1$ is not anymore a pole. Then only the $d-1$ simple poles inside $\Gamma_{-1}$ contribute.

When we evaluate the integral over $u$ we first integrate out the $d-1$ non-trivial poles, that is, the zeros of
\begin{equation}\label{eq43}
R(u,v)=u(1+u)^{d-1}-v(1+v)^{d-1}.
\end{equation}
Let $u_1(v),\ldots,u_{d-1}(v)$ be the solutions of $R(u,v)=0$ different from the trivial one $u_d(v)=v$. Then the kernel for the $d$-spaced initial configuration is as given in (\ref{eq285})-(\ref{eq285b}). A simple case is $d=2$, where the only non-trivial solution of $R(u,v)=0$ is $u=-1-v$.
\end{proofOF}

\section{Asymptotic analysis}\label{SectAsympt}
\subsubsection*{Proof of Theorem~\ref{ThmConvToSasProc}}
To prove Theorem~\ref{ThmConvToSasProc} we consider a conjugate kernel to $K$, i.e., a kernel which gives the same correlation functions. It is given by
\begin{eqnarray}\label{ConjKernel}
& & K^{\rm conj}(n_1,x_1;n_2,x_2)\\
&=&K(n_1,x_1;n_2,x_2)\left(\frac{d}{d-1}\right)^{x_2+d n_2-(x_1+d n_1)} \left(\frac{d^d}{(d-1)^{d-1}}\right)^{n_1-n_2}. \nonumber
\end{eqnarray}
We also use the notation $K^{\rm conj}_0$ for the conjugate of the second part of the kernel $K$, denoted by $K_0$ in (\ref{eq285b}). Notice that the kernel is invariant under simultaneous shifts $x_i\to x_i+d S$ and $n_i\to n_i+S$, $S\in\Z$. For the scaling limit we have to consider, compare with (\ref{eqS1}) and (\ref{eqS2}),
\begin{eqnarray}\label{eqScaling}
x_i&=&\left[-\kappa^2 \frac{2}{d-1}r_i t^{2/3}-\kappa s_i t^{1/3}\right],\nonumber \\
n_i&=&\left[\frac{p(d-1)}{d(d-p)}t+\kappa^2 \frac{2}{d(d-1)}r_i t^{2/3}\right].
\end{eqnarray}

From now on we fix $r_1,\ldots,r_m\in\R$. The dependence on these
constants is not indicated in the following propositions, since it
is irrelevant for the proof of Theorem~\ref{ThmConvToSasProc}.

\begin{figure}[t!]
\begin{center}
\psfrag{-L}[c]{$-L$}
\psfrag{L}[c]{$L$}
\psfrag{e}[c]{$\e t^{2/3}$}
\psfrag{inf}[c]{$\infty$}
\psfrag{s1}[c]{$s_1$}
\psfrag{s2}[l]{$s_2$}
\includegraphics[height=4cm]{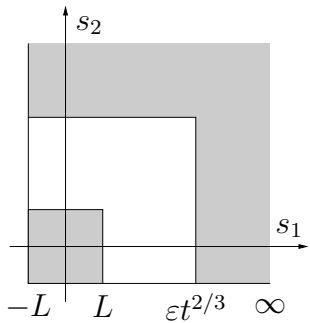}
\caption{The different regimes of $(s_1,s_2)$ for which the kernel is analyzed.}\label{FigBounds}
\end{center}
\end{figure}

To prove convergence of the Fredholm determinant we need to control the behavior of the kernel (\ref{ConjKernel}) as a function of $s_1$ and $s_2$. We do it in three steps corresponding to the three regions indicated in Figure~\ref{FigBounds}.
In Proposition~\ref{PropConvPt} we prove uniform convergence on a bounded set, by controlling the finite-$t$ deviations from the asymptotic term. We thus have the control for \mbox{$(s_1,s_2)\in [-L,L]^2$} for any fixed $L$. Then, in Proposition~\ref{PropModerateDev}, we obtain a bound for $(s_1,s_2)\in [-L,\e t^{2/3}]^2\setminus [-L,L]^2$, for $L$ and $t$ large enough and $\e$ small enough. Finally, in Proposition~\ref{PropLargeDev}, we obtain a bound for
$(s_1,s_2)\in [-L,\infty)^2\setminus [-L,\e t^{2/3}]^2$, for any fixed $\e>0$ and $t$ large.

\begin{prop}[Uniform convergence on compact sets]\label{PropConvPt}
For fixed $r_1$ and $r_2$, the extended kernel has the following limit. Let us fix any $L>0$. Then, with $x_i,n_i$ defined as in (\ref{eqScaling}), the kernel converges uniformly for $(s_1,s_2)\in[-L,L]^2$ as
\begin{eqnarray}
& & \lim_{t\to\infty}K^{\rm conj}(n_1,x_1;n_2,x_2)\kappa t^{1/3}\nonumber \\
& =& -\frac{1}{\sqrt{4\pi (r_2-r_1)}}\exp\left(-\frac{(s_2-s_1)^2}{4 (r_2-r_1)}\right) \Id(r_2>r_1) \\
 & & +\Ai(s_1+s_2+(r_2-r_1)^2) \exp\Big((r_2-r_1)(s_1+s_2)+\frac23(r_2-r_1)^3\Big). \nonumber
\end{eqnarray}
\end{prop}
Proposition~\ref{PropConvPt} leads to a uniform bound on the kernel as in Corollary~\ref{CorBound}. This is obtained using the super-exponential decay of the Airy function, namely, for any $a>0$, $|\Ai(x)|\leq e^{-\tfrac23 x^{3/2}} \leq \cte_a e^{-a x}$ for some $\cte_a$ depending on $a$.
\begin{cor}\label{CorBound}
For any fixed $L>0$ there exists a $t_0=t_0(L)>0$ and a $\cte_L$ independent of $t_0$ s.t.\ for $t>t_0$ the bound
\begin{equation}
\big|K^{\rm conj}(n_1,x_1;n_2,x_2)\kappa t^{1/3}\big| \leq \cte_L
\end{equation}
holds for all $s_1,s_2\in[-L,L]$.
\end{cor}
\begin{prop}[Moderate deviations]\label{PropModerateDev}
For any large enough $L$ there exist an $\e_0=\e_0(L)>0$ and a $t_0=t_0(L)>0$ such that for any
positive $\e<\e_0$ and $t>t_0$ the estimate
\begin{equation}\label{eqLemmaModerate}
|K^{\rm conj}_0(n_1,x_1;n_2,x_2)\kappa t^{1/3}| \leq e^{-(s_1+s_2)}
\end{equation}
holds for $(s_1,s_2)\in [-L,\e t^{2/3}]^2\setminus [-L,L]^2$.
\end{prop}
\begin{prop}[Large deviations]\label{PropLargeDev}
Let $c=(d-1)p/(d-p)$, $\e>0$ as in Proposition~\ref{PropModerateDev} (small enough), and $\tilde s_i=\kappa s_i t^{-2/3}$.\\
(1) If $\tilde s_1\in [\e,c]$ or $\tilde s_2\in [\e,c]$ we have the bound
\begin{equation}
|K^{\rm conj}_0(n_1,x_1;n_2,x_2)| \leq e^{-(s_1+s_2)}
\end{equation}
for $t$ large enough.\\
(2) If $\tilde s_1\in (c,\infty)$ or $\tilde s_2\in (c,\infty)$, then the kernel is zero for $t$ large enough.
\end{prop}
\noindent The last ingredient is an estimate on the binomial part of the kernel.
\begin{prop}\label{PropBound}
For any $s_1,s_2\in \R$ and $r_2-r_1>0$ fixed, the bound
\begin{eqnarray}\label{eq5.66}
& & \kappa t^{1/3} \left(\frac{d}{d-1}\right)^{x_2+d n_2-(x_1+d n_1)} \left(\frac{d^d}{(d-1)^{d-1}}\right)^{n_1-n_2} \binom{x_1-x_2-1}{n_2-n_1-1}\nonumber \\
&\leq & \cte_1 e^{-|s_2-s_1|}
\end{eqnarray}
holds for $t$ large enough and $\cte_1$ independent of $t$.
\end{prop}

\noindent With these results we can now prove Theorem~\ref{ThmConvToSasProc}.
\begin{proofOF}{Theorem~\ref{ThmConvToSasProc}}
From Corollary~\ref{CorBound}, Propositions~\ref{PropModerateDev} and~\ref{PropLargeDev}, it follows that there exist a $t_0>0$ and a $\cte_2$ independent of $t_0$ such that
\begin{equation}\label{eq5.2}
\big|K^{\rm conj}_0(n_1,x_1;n_2,x_2)\kappa t^{1/3}\big| \leq \cte_2 e^{-(s_1\vee 0+s_2\vee 0)}
\end{equation}
for any $t>t_0$. Here $a\vee b=\max\{a,b\}$.

The joint distributions of the rescaled process $X_t$ defined in (\ref{eqRescaled}) are given by a Fredholm determinant with series
\begin{eqnarray}\label{eq5.3}
& &\Pb\Big(\bigcap_{k=1}^m \{X_t(u_k)\leq s_k\}\Big) \nonumber\\
&=&\sum_{n\geq 0}\frac{(-1)^n}{n!}\sum_{i_1,\ldots,i_n=1}^m \int\dx y_1\ldots \dx y_n \prod_{k=1}^n\Id(y_k<\mu u_{i_k} t^{2/3}-\kappa s_{i_k}t^{1/3})\nonumber \\
& &\times \det\left(K^{\rm conj}(n(u_{i_k},t),[y_k];n(u_{i_l},t),[y_l])\right)_{1\leq k,l\leq n}.
\end{eqnarray}
By the change of variables $\sigma_k=(y_k-\mu u_{i_k}t^{2/3})/(-\kappa t^{1/3})$ and a conjugation we obtain
\begin{eqnarray}\label{eq5.4}
& &\hspace{-3em}(\ref{eq5.3})=\sum_{n\geq 0}\sum_{i_1,\ldots,i_n=1}^m \int\dx \sigma_1\ldots \dx \sigma_n \prod_{k=1}^n\Id(\sigma_k>s_{i_k}) \\
& &\hspace{-3em}\times \frac{(-1)^n}{n!}
\det\left(\kappa t^{1/3} K^{\rm conj}(n(u_{i_k},t),[y_k];n(u_{i_l},t),[y_l])\frac{(1+\sigma_l^2)^{i_l}}{(1+\sigma_k^2)^{i_k}}\right)_{1\leq k,l\leq n}. \nonumber
\end{eqnarray}
The term $\frac{(1+\sigma_l^2)^{i_l}}{(1+\sigma_k^2)^{i_k}}$ is the new conjugation, which does not change the determinant.

On $K^{\rm conj}_0$ we use the bound (\ref{eq5.2}). Whenever $i_k>i_l$, we have an additional term coming from the binomial part of the kernel. This contribution is bounded as in Proposition~\ref{PropBound}. Therefore the $(k,l)$ coefficient in the last determinant in (\ref{eq5.4}) is bounded, for $t$ large enough, by
\begin{equation}\label{eq5.5}
\left\{\begin{array}{ll}
\cte_3 (e^{-|\sigma_l-\sigma_k|}+e^{-(\sigma_l\vee 0+\sigma_k\vee 0)})\frac{(1+\sigma_l^2)^{i_l}}{(1+\sigma_k^2)^{i_k}},& i_k>i_l,\\
\cte_3 e^{-(\sigma_l\vee 0+\sigma_k\vee 0)}\frac{(1+\sigma_l^2)^{i_l}}{(1+\sigma_k^2)^{i_k}},& i_k\leq i_l,
\end{array}\right.
\end{equation}
with $\cte_3$ independent of $t$. We also use the bounds, for $i,j\in \{1,\ldots,m\}$,
\begin{equation}
\frac{(1+x^2)^i}{(1+y^2)^{j}}e^{-|x-y|}\leq \frac{1}{1+y^2}\frac{(1+x^2)^i}{(1+y^2)^{i}}e^{-|x-y|}\leq \frac{1}{1+y^2} \cte_4,\quad j>i,
\end{equation}
and
\begin{equation}
\frac{(1+x^2)^i}{(1+y^2)^j}e^{-x\vee 0-y\vee 0}\leq \cte_5 e^{-(x+y)/2}.
\end{equation}
These bounds applied to (\ref{eq5.5}) leads to
\begin{equation}\label{eq5.8}
(\ref{eq5.5})\leq \cte_6\left\{\begin{array}{ll}
(1+\sigma_k^2)^{-1},& i_k>i_l,\\
e^{-(\sigma_k+\sigma_l)/2},& i_k\leq i_l,
\end{array}\right.
\end{equation}
for some $\cte_6$ independent of $t$.

(\ref{eq5.8}) implies that in the determinant in (\ref{eq5.4}) we can single out a product $\prod_{k=1}^n \max\{(1+\sigma_k^2)^{-1},e^{-\sigma_k/2}\}$. Then the entries of the determinant are bounded by $\cte_6$, so that the whole integrand is bounded by
\begin{equation}
\frac{1}{n!} \cte_6^n n^{n/2} \prod_{k=1}^n \max\{(1+\sigma_k^2)^{-1},e^{-\sigma_k/2}\} \Id(\sigma_k>-L)
\end{equation}
where the factor $n^{n/2}$ is the Hadamard bound (the absolute value of a determinant of a $n\times n$ matrix with entries of absolute value not exceeding $1$ is bounded by $n^{n/2}$). Therefore, for $t$ large enough and some $\cte_7>0$ independent of $t$,
\begin{equation}
|(\ref{eq5.3})|\leq \sum_{n\geq 0} \frac{n^{n/2}}{n!}m^n\cte_7^n<\infty
\end{equation}
because $n!$ grows like $(n/e)^n$, which is much stronger than $n^{n/2}$ and any term exponential in $n$.
Therefore for $t$ large enough the integrand can be bounded by a $t$-independent integrable function. By applying dominated convergence, we can exchange the $t\to\infty$ limit with the sums and integrals. The pointwise convergence comes from Proposition~\ref{PropConvPt}, thus
\begin{eqnarray}
& & \lim_{t\to\infty}\Pb\Big(\bigcap_{k=1}^m \{X_t(u_k)\leq s_k\}\Big) \nonumber\\
&=& \sum_{n\geq 0}\frac{(-1)^n}{n!}\sum_{i_1,\ldots,i_n=1}^m \int_{\sigma_k>s_{i_k}}\dx \sigma_1\ldots \dx \sigma_n
\prod_{k=1}^n \det(K_{\rm F_1}(u_{i_k},\sigma_k;u_{i_l},\sigma_l))_{1\leq k,l\leq n}\nonumber \\
&\equiv& \Pb\Big(\bigcap_{k=1}^m \{\Af(u_k) \leq s_k\}\Big).
\end{eqnarray}
Thus $X_t$ converges to $\Af$ as $t\to\infty$ in the sense of finite-dimensional distributions.
\end{proofOF}

\subsubsection*{Proof of the uniform convergence on bounded sets}
\begin{proofOF}{Proposition~\ref{PropConvPt}}
Let us consider the scaling (\ref{eqScaling}). First we determine the contribution coming from the binomial term of (\ref{eq285}), i.e.,
\begin{equation}\label{eq5.6}
\kappa t^{1/3} \left(\frac{d}{d-1}\right)^{x_2+d n_2-(x_1+d n_1)} \left(\frac{d^d}{(d-1)^{d-1}}\right)^{n_1-n_2} \binom{x_1-x_2-1}{n_2-n_1-1}.
\end{equation}
Set
\begin{equation}
a=\frac{2\kappa^2}{d(d-1)}(r_2-r_1)t^{2/3}-1, b=\kappa(s_2-s_1)t^{1/3}+(d-1), \textrm{ and } \e=b/a.
\end{equation}
The binomial term is $\binom{a(d+\e)}{a}$ and $x!=\sqrt{2\pi x}\exp(x\ln(x)-x)(1+\Or(x^{-1}))$. Therefore we have
\begin{equation}\label{eq5.14}
(\ref{eq5.6})=\frac{1+\Or(t^{-1/3})}{\sqrt{4\pi (r_2-r_1)}} e^{-b^2/(2d(d-1)a)}(1+\Or(\e)).
\end{equation}
By reinserting the expressions for $a$ and $b$ we obtain
\begin{equation}\label{eq5.15}
(\ref{eq5.6})=\frac{1}{\sqrt{4\pi (r_2-r_1)}} \exp\left(-\frac{(s_2-s_1)^2}{4(r_2-r_1)}\right) (1+\Or(\e)+\Or(t^{-1/3}))
\end{equation}
with an error uniform for $s_1,s_2\in [-L,L]$. For any fixed
$r_2>r_1$, the error $\Or(\e)$ is also of order $\Or(t^{-1/3})$.
Thus taking the limit $t\to\infty$ we get the first term of the
limit kernel $K_{\rm F_1}$, see (\ref{eqKernelF1}).

Next, we consider the main part of the kernel, i.e.,
(\ref{eq285b}). It is not difficult to see (by looking at the
image of any closed simple loop around $u=-1$  under the map
$u(1+u)^{d-1}$) that the image of $\{u_1(v),\ldots,u_{d-1}(v)\}$ form
closed loops around $-1$ if $v$ is a closed loop around $0$. The
leading term of the kernel is
\begin{equation}\label{eq286}
\frac{1}{2\pi \I} \sum_{i=1}^{d-1} \oint_{\Gamma_0}\dx v \exp\big(t \psi(v)-t \psi(u_i(v))\big)
\end{equation}
with
\begin{equation}
\psi(v)=-\frac{p}{d-p}\ln(v)-\ln(1+pv).
\end{equation}
We can rewrite
\begin{equation}
\Re(\psi(v))=-\frac{p}{d-p}\ln\big(|v| |1/p+v|^{d/p-1}\big)-\ln{p}.
\end{equation}
In Proposition~\ref{Lemma3} we prove that there exists a path $\Gamma_0$
of finite length such that $\Re(\psi(v))-\Re(\psi(u_i(v)))<0$
whenever $u_i(v)\neq v$. Moreover, Proposition~\ref{Lemma3} says
that $\Re(\psi(v))-\Re(\psi(u_i(v)))< 0$ except at the point
$v=-1/d$ for the solution $u_1(v)$ (the one for which
$u_1(-1/d)=-1/d=v$). Therefore the contributions of
$u_2,\ldots,u_{d-1}$ are bounded by $\Or(e^{-\nu_i t}) F(0)$ for
some $\nu_i>0$ and
\begin{equation}
F(0)=\left(\frac{d}{d-1}\right)^{x_1+d n_1-(x_2+d n_2)} \left(\frac{d^d}{(d-1)^{d-1}}\right)^{n_2-n_1}\frac{d}{d-1}.
\end{equation}
The contribution of $u_1$ can be estimated by integrating only on $|v+1/d|\leq \delta$ for some $\delta>0$. The error will be exponentially small in $t$. Thus, let $\Gamma_0^\delta$ denote $\Gamma_0$ restricted to a $\delta$-neighborhood of $-1/d$. Then,
\begin{eqnarray}\label{eq415}
& &K_0(n_1,x_1;n_2,x_2)\kappa t^{1/3}=\Or(e^{-\mu t})F(0)\\
& & + \frac{\kappa t^{1/3}}{2\pi \I}\int_{\Gamma_0^\delta}\dx v
\frac{1+dv}{1+d u_1(v)} \frac{(1+pu_1(v))^t (-u_1(v))^{n_1} (1+v)^{x_2+n_2-2}}{(1+pv)^t (-v)^{n_2} (1+u_1(v))^{x_1+n_1+1}}.\nonumber
\end{eqnarray}
for some $\mu=\mu(\delta)>0$.

From now on we denote $u=u_1$ since the other solutions of $R(u,v)=0$ do not appear anymore.
Close to the critical point we can compute the value of $u$ by the series
\begin{equation}\label{eqU1}
u(v)=-\frac{1}{d}-\big(v+\frac{1}{d}\big)-\frac{2d(d-2)}{3(d-1)}\big(v+\frac{1}{d}\big)^2-\frac{4d^2(d-2)^2}{9(d-1)^2}\big(v+\frac{1}{d}\big)^3+\Or\big(\big(v+\frac{1}{d}\big)^4\big).
\end{equation}
Let $z=v+1/d$ and let the image of $\Gamma_0^\delta$ be
$\eta_\delta$, where $|z|\leq \delta$. Thus we can use Taylor
expansion and the last term of (\ref{eq415}) becomes
\begin{equation}
\frac{\kappa t^{1/3}}{-2\pi \I}\int_{\eta_\delta}\dx z
\exp\big(t f_0(z)+t^{2/3}f_1(z)+t^{2/3}f_2(z)+f_3(z)\big)
\end{equation}
with
\begin{eqnarray}\label{eqU2}
f_0(z)&=&\frac{d^3\kappa^3}{3(d-1)^3}z^3+\Or(z^4),\nonumber \\
f_1(z)&=&-\kappa^2(r_2-r_1)\frac{2}{d(d-1)}\ln((d-1)^{d-1}/d^d)+\frac{\kappa^2(r_2-r_1) d^2}{(d-1)^2}z^2+\Or(z^3),\nonumber \\
f_2(z)&=&-\kappa(s_2-s_1)\ln((d-1)/d)-\kappa \frac{d}{d-1}(s_1+s_2)z+\Or(z^2),\nonumber \\
f_3(z)&=&\ln(d/(d-1))+\Or(z).
\end{eqnarray}
The errors are uniform for $s_i\in [-L,L]$ (the $s$-dependence in the error terms is only in the $f_2$ term). Let $\tilde f_i(z)$ be the expression $f_i(z)$ \emph{without} the error terms. Define $F(z)=\exp(t f_0(z)+t^{2/3}f_1(z)+t^{2/3}f_2(z)+f_3(z))$ and $\tilde F(z)$ similarly. Then
\begin{equation}
\frac{\kappa t^{1/3}}{-2\pi \I}\int_{\eta_\delta}\dx z F(z)=\frac{\kappa t^{1/3}}{-2\pi \I}\int_{\eta_\delta}\dx z \tilde F(z) + \frac{\kappa t^{1/3}}{-2\pi \I}\int_{\eta_\delta}\dx z (F(z)-\tilde F(z)).
\end{equation}
To estimate the last term we use the inequality $|e^x-1|\leq e^{|x|}|x|$. Thus
\begin{eqnarray}
& &\bigg|\frac{\kappa t^{1/3}}{-2\pi \I} \int_{\eta_\delta}\dx z (F(z)-\tilde F(z))\bigg|\\
& \leq & \frac{\kappa t^{1/3}}{2\pi} \int_{\eta_\delta}\dx z |\tilde F(z)| e^{\Or(z^4 t+z^3 t^{2/3}+ z^2 t^{1/3}+z)}\Or(z^4 t+z^3 t^{2/3}+ z^2 t^{1/3}+z)\nonumber \\
&=& \frac{\kappa t^{1/3}}{2\pi} \int_{\eta_\delta}\dx z  |e^{t \tilde f_0(z)(1+\chi_1)+t^{2/3} \tilde f_1(z)(1+\chi_2)+t^{1/3} \tilde f_2(z)(1+\chi_3)}|\nonumber \\
& \times & \Or(z^4 t+z^3 t^{2/3}+ z^2 t^{1/3}+z)\nonumber
\end{eqnarray}
for some $\chi_1,\chi_2,\chi_3$ which can be made as small as desired by choosing $\delta$ small enough. We take as integration path $\eta_\delta=\{e^{-\I\pi\sgn(w)/3}|w|,w\in [-\delta,\delta]\}$ which is close to the steepest descent path when $w\to 0$.

At the integration boundaries, $w=\pm\delta$, the leading term is $$\exp\left(-\frac{d^3\kappa^3}{3(d-1)^3}\delta^3(1+\chi_1)t\right),$$ thus the integral remains bounded as $t\to\infty$ and, by the change of variable $w t^{1/3}=q$ we see that the error term is $\Or(t^{-1/3})F(0)$.

The final step is to compute $\frac{\kappa t^{1/3}}{-2\pi \I}\int_{\eta_\delta}\dx z \tilde F(z)$. Extending $\delta$ to $\infty$ we collect only an error of order $\Or(e^{-\mu t})F(0)$ with $0<\mu\sim \delta^3$. This leads to the integration on the path $\eta_\infty=\{e^{-\I\pi\sgn(w)/3}|w|,w\in \R\}$. Thus
\begin{eqnarray}
& &\frac{\kappa t^{1/3}}{-2\pi \I}\int_{\eta_\infty}\dx z \tilde F(z) = F(0) \frac{2 \kappa t^{1/3}e^{-\I\pi/3}}{-2\pi \I}\int_{\R_+}\dx w \exp\left(-\frac{d^3\kappa^3}{3(d-1)^3}w^3t\right) \nonumber \\
&\times & \exp\left(\frac{\kappa^2(r_2-r_1) d^2}{(d-1)^2}e^{-\I 2\pi/3}w^2t^{2/3}\right) \exp\left(-\kappa\frac{d}{d-1}(s_1+s_2)e^{-\I \pi/3}wt^{1/3}\right).\nonumber
\end{eqnarray}
The change of variable $q=w\kappa e^{-\I\pi/3}d/(d-1)$ gives
\begin{eqnarray}
& &\frac{\kappa t^{1/3}}{-2\pi \I}\int_{\eta_\infty}\dx z \tilde F(z) = G(0) \frac{1}{-2\pi \I} \int_{\eta_\infty}\dx q e^{q^3/3} e^{(r_2-r_1)q^2} e^{-(s_1+s_2) q}
\end{eqnarray}
with $G(0)=F(0)(d-1)/d$. By considering the conjugate kernel $K^{\rm conj}_0$ instead of $K_0$, the term $G(0)$ cancels.

Finally we use an Airy function representation
\begin{equation}
\frac{1}{-2\pi \I}\int_{\eta_\infty}\dx v e^{v^3/3+av^2+bv}=\Ai(a^2-b)\exp(2a^3/3-ab)
\end{equation}
to obtain the final result
\begin{equation}\label{eqCVpt}
\lim_{t\to\infty}\frac{\kappa t^{1/3}}{G(0)} K_0(x_1,n_1;x_2,n_2) = \Ai(s_1+s_2+(r_2-r_1)^2)e^{2(r_2-r_1)^3/3+(s_1+s_2)(r_2-r_1)}
\end{equation}
uniformly for $s_1,s_2\in[-L,L]$.
\end{proofOF}
The goal of the following sequence of lemmas is Proposition~\ref{Lemma3}
used in the proof of Proposition~\ref{PropConvPt} above.

\begin{lem}\label{Lemma1}
Define the path
\begin{equation}\label{eqgamma1}
\gamma_1=\{-1+\sin(\phi(d-1))e^{i\phi}/\sin(\phi d),\phi\in[0,\pi/d)\}.
\end{equation}
On $\gamma_1$, $v(1+v)^{d-1}\in\R$. Let $u_i$, $i=1,\ldots,d$, be the $d$ solutions of the equation
\begin{equation}
R(u,v)\equiv u(1+u)^{d-1}-v(1+v)^{d-1}=0.
\end{equation}
Then, for all $v\in\gamma_1$ and $u_i\not\in\{v,\bar{v}\}$, we
have
\begin{equation}
\left|u_i(1/p+u_i)^{d/p-1}\right|\neq \left|v(1/p+v)^{d/p-1}\right|.
\end{equation}
\end{lem}

\begin{lem}\label{Lemma2}
On $\gamma_1\setminus\{-1/d\}$ the solutions $u(v)$ of $R(u,v)=0$ are simple zeros.
\end{lem}

\begin{lem}\label{Lemma2b}
Assume that $a,b\in\mathbb{R}$ satisfy either\\
(1) $b<a<0$ or\\
(2) $a<b$ and $a(a-1)<b(b-1)$. Then, for $\mu>1$, we have the strict inequality
\begin{equation}
\mu^a-\mu^b<(\mu-1)(a-b).
\end{equation}
\end{lem}

\begin{prop}\label{Lemma3}
Let $u_1(v)$ be the solution of $R(u,v)=0$ such that
$u_1(-1/d)=-1/d$, and $u_2(v),\ldots,u_{d-1}(v)$ the other non-trivial $d-2$
solutions. Then, there exists a path $\Gamma_0$ encircling the
origin, passing through $v=-1/d$ such that
\begin{equation}
\Re(\psi(v))-\Re(\psi(u_i(v)))<0
\end{equation}
except for the solution $u_1(v)$ at the point $v=-1/d$.
\end{prop}

\begin{proofOF}{Lemma~\ref{Lemma1}}
For $d=2$ nothing has to be shown because there are only two solutions, $u=v$ and $u=\bar{v}$. Thus we consider $d\geq 3$. Since for $v\in \gamma_1$, $v(1+v)^{d-1}\in\R$, we have to prove that the system of equations
\begin{equation}\label{eqSyst}
\begin{array}{rcl}
|u| |1+u|^{d-1}&=&|v| |1+v|^{d-1}\\
|u| |1/p+u|^{d/p-1}&=&|v| |1/p+v|^{d/p-1}
\end{array}
\end{equation}
has only the trivial solutions $u=v$ and $u=\bar{v}$. If (\ref{eqSyst}) has a solution $u$,
 then by symmetry $\bar{u}$ is also a solution.
Thus in the rest of the proof we restrict ourselves to the
upper-half plane.

(\ref{eqSyst}) is an equation involving only distances of $u$ and $v$ from the points $0$, $-1$, and $-1/p$. Let us choose any $v$ and set
\begin{equation}
a=|v|,\quad b=|1+v|,\quad c=|1/p+v|.
\end{equation}
Similarly we choose a $u\neq v$ and set
\begin{equation}
\tilde a=|u|,\quad \tilde b=|1+u|,\quad \tilde c=|1/p+u|.
\end{equation}
Then (\ref{eqSyst}) writes
\begin{equation}\label{eq292}
\tilde a \tilde b^{d-1}= a b^{d-1},\quad \tilde a \tilde c^{d/p-1}= a c^{d/p-1},
\end{equation}
that is,
\begin{equation}\label{eq293}
\tilde a=a\mu,\quad \tilde b = b/\mu^{1/(d-1)},\quad \tilde c=c/\mu^{p/(d-p)}.
\end{equation}

For $\mu=1$, we have $u=v$. Thus consider $\mu\neq 1$. Given $a,b$ such that the circle centered at $0$ of radius $a$ intersects the circle centered at $-1$ of radius $b$, the position of $v$ is uniquely determined, see Figure~\ref{FigSteepest}.
\begin{figure}[t!]
\begin{center}
\psfrag{a}[c]{$a$}
\psfrag{b}[c]{$b$}
\psfrag{c}[c]{$c$}
\psfrag{beta}[l]{$\pi-\beta$}
\psfrag{phi}[l]{$\phi$}
\psfrag{0}[c]{$0$}
\psfrag{1}[c]{$-1$}
\psfrag{p}[c]{$-1/p$}
\includegraphics[bb=450 0 0 200,clip,height=4cm]{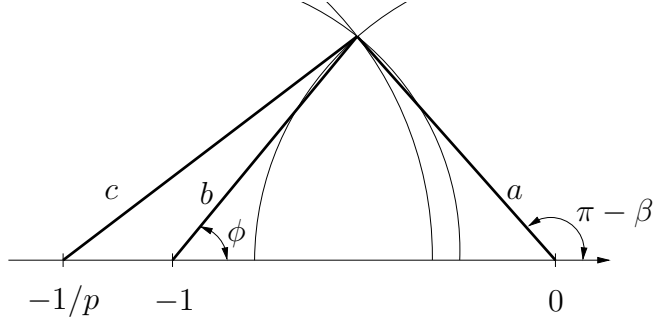}
\caption{Geometric representation of (\ref{eqSyst}).}\label{FigSteepest}
\end{center}
\end{figure}
Thus this determine $c$ too, i.e., there exists a function $f$ to be determined such that $c^2=f(a,b)$. Define the function
\begin{equation}
F(\mu,a,b)\equiv f(\mu a, b/\mu^{1/(d-1)})-f(a,b)/\mu^{2p/(d-p)}.
\end{equation}
Then since also $\tilde c^2=f(\tilde a,\tilde b)$, the scaling
relations (\ref{eq293}) imply that the Lemma will be proven once
we show that $F(\mu,a,b)=0$ only at $\mu=1$ for $a,b$ such that
$v\in\gamma_1$.

To determine the function $f$ we use elementary trigonometry. Let $\pi-\beta$ be the argument of $v$. Then
\begin{eqnarray}
b^2&=&a^2+1-2a\cos(\beta),\nonumber \\
c^2&=&a^2+p^{-2}-2ap^{-1}\cos(\beta).
\end{eqnarray}
From this it follows
\begin{equation}
c^2\equiv f(a,b)=\frac{1-p}{p^2}+\frac{b^2}{p}-\frac{1-p}{p}a^2.
\end{equation}
$v\in\gamma_1$, is parametrized by an angle $\phi$ and we have
\begin{equation}
a=\frac{\sin(\phi)}{\sin(\phi d)},\quad b=\frac{\sin(\phi(d-1))}{\sin(\phi d)}.
\end{equation}
Therefore
\begin{eqnarray}
F(\mu,a,b)&=& \frac{1-p}{p^2}\left(1-\mu^{-2p/(d-p)}\right)
 +\frac{1-p}{p}\frac{\sin^2(\phi)}{\sin^2(\phi d)}\left(\mu^{-2p/(d-p)}-\mu^2\right)
\nonumber \\
&&+ \frac{\sin^2(\phi(d-1))}{p\sin^2(\phi d)}\left(\mu^{-2/(d-1)}-\mu^{-2p/(d-p)}\right)
\end{eqnarray}
Define the function
\begin{eqnarray}
G(\mu)&=&\frac{1-p}{p^2}\left(1-\mu^{-2p/(d-p)}\right)
+\frac{1-p}{pd^2}\left(\mu^{-2p/(d-p)}-\mu^2\right)\nonumber \\
 & & +\frac{(d-1)^2}{p d^2}\left(\mu^{-2/(d-1)}-\mu^{-2p/(d-p)}\right)
\end{eqnarray}
Using the properties
\begin{enumerate}
\item[1)] $\frac{\sin^2(\phi)}{\sin^2(\phi d)}$ and $\frac{\sin^2(\phi(d-1))}{\sin^2(\phi d)}$ are positive and increasing function in $\phi$ for \mbox{$\phi\in[0,\pi/d)$},
\item[2)] $\mu^{-2p/(d-p)}-\mu^2>0$ for $\mu\in (0,1)$ and $\mu^{-2p/(d-p)}-\mu^2<0$ for $\mu>1$,
\item[3)] $\mu^{-2/(d-1)}-\mu^{-2p/(d-p)}>0$ for $\mu\in (0,1)$ and $\mu^{-2/(d-1)}-\mu^{-2p/(d-p)}<0$ for $\mu>1$,
\end{enumerate}
we obtain that
\begin{eqnarray}
F(\mu,a,b)\geq G(\mu),&\textrm{ for }&\mu\in (0,1), \nonumber \\
F(\mu,a,b)\leq G(\mu),&\textrm{ for }&\mu>1.
\end{eqnarray}
Therefore we have to prove that $G(\mu)>0$ for $\mu\in (0,1)$ and $G(\mu)<0$ for $\mu>1$. This follows from the fact that
\begin{enumerate}
\item[1)] $G(1)=0$ (trivial verification) and
\item[2)] $G'(\mu)<0$ for all $\mu\neq 1$ (to be proven below).
\end{enumerate}
$G'(\mu)$ is given by
\begin{equation}
H(\mu)\equiv \frac{pd^2}{2} \mu G'(\mu)= \frac{d-p}{\mu^{2p/(d-p)}}-\frac{d-1}{\mu^{2/(d-1)}}-(1-p)\mu^2.
\end{equation}

Consider first $\mu>1$. We can rewrite
\begin{equation}
H(\mu)=(d-1)\left(\mu^{-2p/(d-p)}-\mu^{-2/(d-1)}\right)+(1-p)\left(\mu^{-2p/(d-p)}-\mu^2\right).
\end{equation}
By Lemma~\ref{Lemma2b} if follows, for $\mu>1$, that
\begin{equation}\label{eq4.43}
A(\mu)=\mu^{-2p/(d-p)}-\mu^{-2/(d-1)} < \left(\frac{2}{d-1}-\frac{2p}{d-p}\right)(\mu-1).
\end{equation}
Moreover, for $d\geq 3$, by Lemma~\ref{Lemma2b} we have that
\begin{equation}
\mu^{-2p/(d-p)}-\mu^2 < -\frac{2d}{d-p}(\mu-1)
\end{equation}
holds for $\mu>1$. Using these bounds we obtain $H(\mu)<0$ for $\mu>1$, from which $G'(\mu)<0$ for $\mu>1$ too.

Now consider the case $\mu\in (0,1)$. We get
\begin{equation}
\mu H'(\mu)/2=\left(\mu^{-2/(d-1)}-\mu^{-2p/(d-p)}\right)+(1-p)\left(\mu^{-2p/(d-p)}-\mu^2\right)>0
\end{equation}
because both terms are strictly positive for $\mu\in (0,1)$. Since $H(1)=0$, we then get $H(\mu)<0$ for $\mu\in (0,1)$. Thus $G'(\mu)<0$ for $\mu\in(0,1)$ too, and this finishes the proof of Lemma~\ref{Lemma1}.
\end{proofOF}

\begin{proofOF}{Lemma~\ref{Lemma2}}
On $\gamma_1$ we have $v(1+v)^{d-1}\in\R$. By symmetry if $u(v)$ is a solution of $R(u,v)=0$, then also $\bar{u}(v)$ is a solution of the same equation. Thus we have to check that $u=v$ is a simple zero for $v\in\gamma_1\setminus\{-1/d\}$. We have $R(u,v)=\prod_{k=1}^d(v-u_k(v))$. If $u=v$ would be a double solution, then the Taylor expansion of $R(u,v)$ at $u=v$ would have the form $\sum_{k\geq 1}a_k(v) (v-u)^k$ with $a_1(v)=0$. However, explicit computations gives
\begin{equation}
R(u,v)=-v(1+v)^{d-1}\left(\frac{1}{v}+\frac{d-1}{v+1}\right)(v-u)+a_2(v)(v-u)^2+\cdots.
\end{equation}
We see that the only $v\in\gamma_1$ such that $a_1(v)=0$ is $v=-1/d$. Thus Lemma~\ref{Lemma2} is proven.
\end{proofOF}

\begin{proofOF}{Lemma~\ref{Lemma2b}}
Consider the function $f(\mu)=\mu^a-\mu^b$. Then its second derivative is
\begin{equation}
f''(\mu)=\big(a(a-1)\mu^a-b(b-1)\mu^b\big) \mu^{-2}.
\end{equation}
For $b>a$ and $\mu>1$ we then have $f''(\mu)<\big(a(a-1)-b(b-1)\big)\mu^{a-2}< 0$ as soon as $a(a-1)-b(b-1)<0$. In this case $f(\mu)$ is strictly concave for $\mu>1$ and case (2) of Lemma~\ref{Lemma2b} is proven.\\
For $b<a<0$, $f(\mu)$ is not anymore concave for all $\mu>1$. Let us compute the zero of $f'(\mu)$, $\mu_1$, and the zero of $f''(\mu)$, $\mu_2$. We get
\begin{eqnarray}
\mu_1&=&\exp\bigg(\frac{\ln(b/a)}{a-b}\bigg)>1,\\ \mu_2&=&\exp\bigg(\frac{\ln(b/a)}{a-b}\bigg)\exp\bigg(\frac{\ln((b-1)/(a-1))}{a-b}\bigg)>\mu_1. \nonumber
\end{eqnarray}
This means that $f(\mu)$ is strictly concave as long as it increases and at $\mu=1$ it is still increasing. Thus the bound for case (1) follows.
\end{proofOF}

\begin{proofOF}{Proposition~\ref{Lemma3}}
We first define $\Gamma_0$ close to infinity and the critical point. Secondly, using Lemma~\ref{Lemma1} we prove that we can complete the path satisfying Proposition~\ref{Lemma3}. The way we do it is illustrated in Figure~\ref{FigPathGamma0}.
\begin{figure}[t!]
\begin{center}
\psfrag{-1}[r]{$-1$}
\psfrag{pi/d}[l]{$\pi/d$}
\psfrag{rho}[l]{$\rho$}
\psfrag{g1}[l]{$\gamma_1$}
\psfrag{G0}[l]{$\Gamma_0$}
\includegraphics[height=5cm]{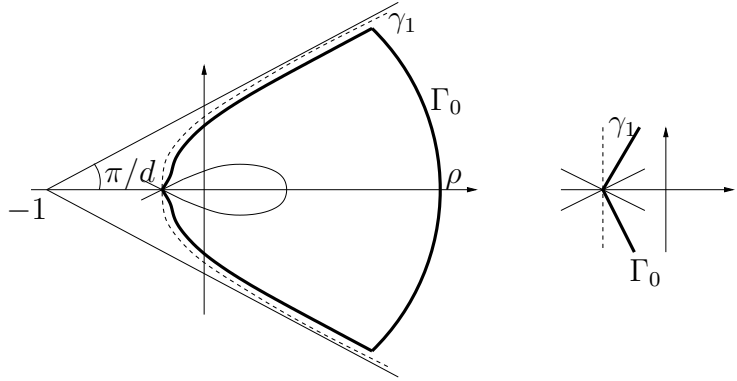}
\caption{Illustration of the path $\Gamma_0$ (the bold line) used in the asymptotic analysis. The dashed line is the $\gamma_1$. On the right the structure close to the critical point is shown.}\label{FigPathGamma0}
\end{center}
\end{figure}

First we consider the infinity, i.e., $v=\rho e^{\I\varphi}$ for $\rho\gg 1$, and $\varphi\in (-\pi/d,\pi/d)$. Then the $d-1$ non-trivial solutions $u_k(v)$, $k=1,\ldots,d-1$, (different from the $u=v$ one) of $R(u,v)=0$ can be expanded in power of $\rho^{-1}$. Let us set
\begin{equation}
u_k=\rho e^{\I\varphi}e^{-\I 2\pi k/d}(1+\alpha_k/\rho+\Or(\rho^{-2})).
\end{equation}
Simple calculations lead to
\begin{equation}
\alpha_k=\frac{d-1}{d}\big(1-(-1)^{2(d-k)(d-1)/d}\big)e^{-\I\varphi},
\end{equation}
from which
\begin{equation}
\Re(\psi(v))-\Re(\psi(u_k(v)))=-\frac{(1-p)d}{\rho p (d-p)}
\Re\big(e^{\I\varphi}-e^{\I\varphi}e^{-\I 2\pi k/d}\big)+\Or(\rho^{-2}).
\end{equation}
Since for $k=1,\ldots,d-1$ and $\varphi\in (-\pi/d,\pi/d)$, $\Re\big(e^{\I\varphi}-e^{\I\varphi}e^{-\I 2\pi k/d}\big)>0$, it follows that for $\rho$ large enough, $\Re(\psi(v))-\Re(\psi(u_k(v)))<0$. This strict inequality holds also for $\varphi=\pm \pi/d$ for $k=2,\ldots,d-1$.
Therefore we can choose a part of $\Gamma_0$ to be parametrized by $v=\rho e^{\I\varphi}$ for all $\varphi\in(-\pi/d,\pi/d)$.

Next we look close to the critical point. The above strict inequality holds also for $\varphi=\pm \pi/d$ for $k=2,\ldots,d-1$, thus also on the line $\gamma_1$. By continuity and Lemma~\ref{Lemma1}, $\Re(\psi(v))-\Re(\psi(u_k(v)))$ does not change its sign along $\gamma_1$, thus we can follow the path $\gamma_1$ to come back to the critical point, with the result
\begin{equation}
\Re(\psi(-1/d))-\Re(\psi(u_k(-1/d)))<0, \quad k=2,\ldots,d-1.
\end{equation}
By continuity, $\Re(\psi(v))-\Re(\psi(u_k(v)))<0$ in a small enough neighborhood of $-1/d$. Thus any path leaving from $-1/d$ is good, at least locally, for the $d-2$ solutions. Next, we consider the solution $u_1(v)$ close to the critical point. It is the only non-trivial solution that $u_1(-1/d)=-1/d$ (see Lemma~\ref{Lemma2}). By Taylor expansion we obtain (\ref{eqU1}), which yields
\begin{equation}
\psi(v)-\psi(u_1(v))=\frac{d^3\kappa^3}{3(d-1)^3}z^3+\Or(z^4),\quad z=v+1/d,
\end{equation}
see (\ref{eqU2}). Thus in a neighborhood of the critical point there are three lines along which $\Re(\psi(v))-\Re(\psi(u_1(v)))=0$. These lines are locally given by
\begin{equation}
v=-\frac{1}{d}+\I x+\Or(x^2),\quad v=-\frac{1}{d}+e^{\I\pi/6}x+\Or(x^2),\quad v=-\frac{1}{d}+e^{\I 5\pi/6}x+\Or(x^2),
\end{equation}
for $x$ real, see Figure~\ref{FigPathGamma0}.
Therefore we can choose the path $\Gamma_0$ close to the critical point to be any path of the form
\begin{equation}
v=-\frac{1}{d} + e^{\I\theta} x+\Or(x^2),\quad \theta \in (-\pi/2,-\pi/6)\cup(\pi/6,\pi/2).
\end{equation}
The steepest descent path leaves the critical point with an angle $\pm\pi/3$.

The final step is to see that we can join the part of the path $\Gamma_0$ close to the critical point and the one far away (at large enough distance $\rho$ from the origin) by going close to $\gamma_1$. Assume that we can not do that. Then somewhere along the way we must hit a point where $\Re(\psi(v))=\Re(\psi(u_i(v))$ with $u_i(v)\neq v$. Let us call points $v$ with this property ``bad''. Thus, we know that there are bad points at arbitrarily small distances to $\gamma_1$, and also these points have bounded absolute values because the neighborhood of infinity is completely controlled.

Therefore, the sequence of bad points necessarily has a limit point on $\gamma_1$ which is not $-1/d$ (because near $-1/d$ everything is controlled, too). But because of Lemma~\ref{Lemma1}, this implies that bad points near $\gamma_1$ can come only from the root $u_1(v)$ which is equal to $\bar{v}$ on $\gamma_1$. Thus we have to prove that the root $u_1$ does not cause bad points. For this, the neighborhood of $\gamma_1$ away from the critical point can be parametrized as
\begin{equation}
v(\varphi,\theta)=\Big(-1+\frac{\sin((d-1)\varphi)}{\sin(d\varphi)}e^{\I\varphi}\Big)e^{-\I\theta}.
\end{equation}
For $\varphi\in(0,\pi/d)$, $1\gg \theta>0$ means that the point has increased real and decreases imaginary part with respect to $\theta=0$. We can compute explicitly $u_1(v)$ in series of $\theta$, with the result
\begin{equation}
u_1(\varphi,\theta)=v(-\varphi,\theta)(1+\alpha\theta+\Or(\theta^2))
\end{equation}
with
\begin{equation}
\alpha=-\I\frac{1+v(\varphi,0)d}{1+v(\varphi,0)}\frac{1+v(-\varphi,0)}{1+v(-\varphi,0)d}.
\end{equation}
Let $v(\varphi,0)=x+\I y$. Then explicit computations lead to
\begin{eqnarray}
& &\!\!\frac{d}{d\theta}\big(\Re\psi(v)-\Re\psi(u(v))\big)\Big|_{\theta=0}= \frac{py}{(d-p)((1+x)^2+y^2)((1+px)^2+p^2y^2)}\nonumber \\
& &\!\!\times \big((p+2pd-3d)(y^2+x^2)+(3p-2-d)+(4d+2pd-2-4d)x\big) \nonumber
\end{eqnarray}
The denominator is always positive. Consider $y>0$ (i.e., away from the critical point). Then
$p+2pd-3d=p(1-p)+3(p-1)d<0$ for $p\in(0,1)$ and $d\geq 2$. Thus the term in $y^2$ is always strictly negative. To analyze the contribution in $x$, we set $x=-1/d+z$, and on $\gamma_1\setminus\{-1/d\}$ we have $z>0$. As a function of $z$, the remainder of the numerator becomes
\begin{eqnarray}
& &-\frac{(d-1)^2(d-p)}{d^2}-\frac{2(d-1)(p(d-1)-2d(1-p))}{d}z\nonumber \\
& &-(p(d-1)+3d(1-p))z^2<0.
\end{eqnarray}
Therefore, in a right-neighborhood of $\gamma_1$, $\Re(\psi(v))-\Re(\psi(u_1(v)))<0$. Thus $u_1(v)$ does not generate bad points in a right-neighborhood of $\gamma_1$ (excluding $\gamma_1$).
\end{proofOF}

In the above construction of the contour $\Gamma_0$ we ignored the requirement that the contour is not allowed to contain any additional poles, see Theorem~\ref{ThmKernel}. These additional poles appear as $u_i(v)=-1/d$. In particular, if $u(1+u)^{d-1}$ is real, then also $v(1+v)^{d-1}$ is real. This happens on the $d-1$ branches, one being $\gamma_1$ from (\ref{eqgamma1}), the others are $d-2$ branches originate at $-1$ and going to infinity in the directions $e^{\pm \I k \pi/d}$, $k=2,\ldots,d-1$. These branches do not intersect with $\gamma_1$ because of Lemma~\ref{Lemma2}, thus do not intersect the interior of $\Gamma_0$. Therefore the path $\Gamma_0$ fulfills all the requirements of Theorem~\ref{ThmKernel}.

\subsubsection*{Proof of the bound for moderate deviations}
\begin{proofOF}{Proposition~\ref{PropModerateDev}}
In this proof we set $\tilde s_i=\kappa s_i t^{-2/3}$. The $\e>0$ is still to be chosen.
We can set it as small as desired, but of course independent of $t$. For $\e>0$ small enough, the path $\Gamma_0$ can be chosen to be equal to the path used in the proof of Proposition~\ref{Lemma3} and illustrated in Figure~\ref{FigPathGamma0}, except for a deformation close to the critical point.

We set
\begin{equation}
\psi_{\e,i}(v)=-\frac{p}{d-p}\ln(v/u_i(v))-\ln\left(\frac{1+pv}{1+p u_i(v)}\right)+\tilde s_2 \ln(1+v)-\tilde s_1\ln(1+u_i(v)).
\end{equation}

First consider the $\e=0$ situation. Let $u_i(v)$ be the solutions of $R(u,v)=0$ as in Proposition~\ref{Lemma3}. It follows from Proposition~\ref{Lemma3} that there exists a $\delta>0$ such that $\Re(\psi_{\e,1}(v)) \leq -\delta$ for all $v\in \Gamma_0\setminus\{|v+1/d|\leq \cte_8 \delta^{2/3}\}$ for some $\cte_8>0$ small enough. Next consider the situation we want to analyze, i.e., $\e>0$ small. Since $\Gamma_0$ remains on a bounded region, we can choose $\e$ with $0<\e\ll \delta^{4/3}$ such that $\Re(\psi_{\e,1}(v))\leq \delta/2$ for $v$ as above.
Similarly, for $i=2,\ldots,d-1$, $\Re(\psi_{\e,i}(v))\leq \delta/2$ for all $v\in\Gamma_0$, since the only solution $u_i(v)$ such that $\Re(\psi_{0,i}(-1/d))=0$ is $u_1(v)$.

Thus we choose $\Gamma_0$ as in Proposition~\ref{Lemma3} for $\{|v+1/d|\geq c \delta^{2/3}\}$, and modify only the path in the region $\{|v+1/d|\leq c \delta^{2/3}\}$. The contribution coming from the unmodified path is then bounded by $\Or(e^{-\delta t/2})<\Or(e^{-2 \e^{3/4}t})<\Or(e^{-2 \e^{3/2}t})$. Now, note that $s_i^{3/2}\leq \e^{3/2}t$, thus the previous contribution is bounded by $\Or(e^{-(s_1+s_2)^{3/2}})$.

Next we have to consider the neighborhood of $v=-1/d$ and see how the positions of the critical points depends on $\e$. We will then be able to choose the appropriate path locally. We have already computed $u_1(v)$ in the neighborhood of $v=-1/d$ in (\ref{eqU1}). The doubly critical point at $v=-1/d$ is separates now into two critical points $v_\pm$ on the real axis. They are the stationary points of $\Re(\psi_{\e,1}(v))-\Re(\psi_{\e,1}(-1/d))$ and can be computed in series of $\sqrt{\tilde s_1+\tilde s_2}$. Explicitly we have
\begin{equation}\label{eq465}
\Re(\psi_{\e,1}(v))-\Re(\psi_{\e,1}(-1/d))=\frac{d^3 \kappa^3}{3(d-1)^3}z^3(1+\Or(z))-\frac{(\tilde s_1+\tilde s_2)d}{d-1}z(1+\Or(z))
\end{equation}
with $z=v+1/d$. From (\ref{eq465}) it follows that
\begin{equation}
v_\pm=-\frac{1}{d}\pm \sqrt{\tilde s_1+\tilde s_2} \frac{(d-1)}{d \kappa^{3/2}}(1+\Or(\sqrt{\tilde s_1+\tilde s_2})).
\end{equation}
$\Re(\psi_{\e,1}(v))$ is minimal at $v_+$. Therefore we can choose the path $\Gamma_0$ to pass by $v_+$. Since we need just a bound, we can disregard the error term and set the path $\Gamma_0$ to pass by
\begin{equation}
v_+^{0}=-\frac{1}{d}\pm \sqrt{\tilde s_1+\tilde s_2} \frac{(d-1)}{d \kappa^{3/2}}.
\end{equation}
The steepest descent path leaves $v_+^0$ along the imaginary direction, see Figure~\ref{FigModerateDev}, thus the final choice of $\Gamma_0$ in the $\delta$-neighborhood of $-1/d$ is simply $\eta_0=\{v_+^0 +\I y,|y|\leq \sqrt{\tilde s_1+\tilde s_2} \frac{(d-1) \sqrt{3}}{d \kappa^{3/2}}\}$ joint with $\eta_1$ and $\bar{\eta}_1$.
\begin{figure}[t!]
\begin{center}
\psfrag{vc}[c]{$-\frac1d$}
\psfrag{vp}[l]{$v_+^0$}
\psfrag{g0}[l]{$\eta_0$}
\psfrag{g1}[l]{$\eta_1$}
\psfrag{g1b}[l]{$\bar{\eta}_1$}
\includegraphics[height=5cm]{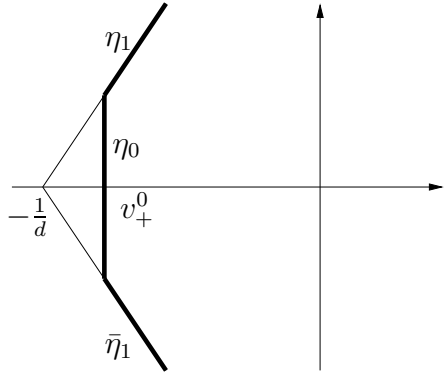}
\caption{Local choice of the path $\Gamma_0$ (the bold line) used in the asymptotic analysis for moderate deviations.}\label{FigModerateDev}
\end{center}
\end{figure}

We change the variable by setting $v=-1/d+z$, and denote $z_+=v_+^0 +1/d$. Then the contribution coming from the integral over $\Gamma_0$ in the $\delta$-neighborhood of $-1/d$ is
\begin{equation}\label{eq5.61}
\frac{\kappa t^{1/3}}{-2\pi \I}\int_{\bar{\eta}_1\vee \eta_0\vee\eta_1}\dx z \exp\big(F_t(z)\big)
\end{equation}
where $F_t(z)=t f_0(z)+t^{2/3}f_1(z)+t^{1/3}f_2(z)+f_3(z)$, with
\begin{eqnarray}\label{eqU3}
f_0(z)&=&\!\frac{d^3 \kappa^3}{3(d-1)^3}z^3(1+\Or(z))-\frac{(\tilde s_1+\tilde s_2)d}{d-1}z(1+\Or(z)),\nonumber \\
f_1(z)&=&\!-\kappa^2(r_2-r_1)\frac{2}{d(d-1)}\ln\left(\frac{(d-1)^{d-1}}{d^d}\right)+\frac{\kappa^2(r_2-r_1) d^2}{(d-1)^2}z^2+\Or(z^3),\nonumber \\
f_2(z)&=&\!-\kappa(s_2-s_1)\ln((d-1)/d),\nonumber \\
f_3(z)&=&\!\ln(d/(d-1))+\Or(z).
\end{eqnarray}
This is like (\ref{eqU2}) except that now the $s_i$ are in the leading term $f_0$ since they can be very large, namely of order $\Or(t^{2/3})$.

On $\eta_0$, $z=\sqrt{\tilde s_1+\tilde s_2} \frac{(d-1)}{d \kappa^{3/2}}(1+\I w)$ with $w\in (-\sqrt{3},\sqrt{3})$. It is not difficult to check that on $\eta_0$
\begin{eqnarray}
\Re(t f_0(z))&\leq& -\frac23 (s_1+s_2)^{3/2}(1+\Or(\sqrt{\tilde s_1+\tilde s_2})),\nonumber\\
\Re(t^{2/3} f_1(z))&\leq & t^{2/3} f_1(0)+(s_1+s_2)(1+\Or(\sqrt{\tilde s_1+\tilde s_2})),\nonumber\\
\Re(f_3(z))&\leq & f_3(0)+\Or(\sqrt{\tilde s_1+\tilde s_2}).
\end{eqnarray}
For $L$ large enough, the linear term in $s_1+s_2$ is controlled by the $-\frac23 (s_1+s_2)^{3/2}$ term. Thus for $\e$ small enough, the contribution (\ref{eq5.61}) coming from $\eta_0$ can be bounded as
\begin{equation}
|\eta_0| \kappa t^{1/3} e^{F_t(0)} e^{-\frac13 (s_1+s_2)^{3/2}}.
\end{equation}
But $|\eta_0|=2t^{-1/3}\sqrt{s_1+s_2} \frac{(d-1) \sqrt{3}}{d \kappa}$, therefore the $t^{1/3}$ factor simplifies and the contribution (\ref{eq5.61}) coming from $\eta_0$ is bounded by $\cte_9 e^{F_t(0)} e^{-\frac13 (s_1+s_2)^{3/2}}$ for some finite $\cte_9>0$.

The final step is to bound the contribution coming from the integration on $\eta_1$ (the same bound holds for the $\bar{\eta}_1$ contribution). It is parametrized by
$z=e^{\I\pi/3}y$, $y\in [z_+/2,\delta]$. The term $t^{2/3} f_1(z)$ is dominated, for large $L$, by the cubic term in $z$ coming from the term $t f_0(z)$. In fact, while $y$ increases the cubic term becomes more and more dominating with respect to the quadratic term in $z$ and from the analysis on $\eta_0$, the domination occurs already at $y=z_+/2$.

The linear term in $z$ of $f_3(z)$ is dominated by the linear term in $z$ coming from $t f_0(z)$, again for $L$ large enough. Therefore for $L\gg 1$ and $\delta \ll 1$,
\begin{eqnarray}
\Re(F_t(z)) & \leq & F_t(0) -t\frac{d^3\kappa^3}{6(d-1)^3}y^3-t\frac{(\tilde s_1+\tilde s_2)d}{4(d-1)}y \nonumber \\
&\leq &F_t(0) -t\frac{d^3\kappa^3}{6(d-1)^3}y^3-t\frac{(\tilde s_1+\tilde s_2)^{3/2}}{8\kappa^{3/2}}
\end{eqnarray}
where the last inequality follows from $y\geq z_+/2$. Thus
\begin{equation}
\left|\frac{\kappa t^{1/3}}{-2\pi \I}\int_{\eta_1}\dx z e^{F_t(z)}\right| \leq \cte_{10} e^{F_t(0)} e^{-\frac18 (s_1+s_2)^{3/2}} \int_0^\infty\dx y t^{1/3} e^{-t\frac{d^3\kappa^3}{6(d-1)^3}y^3}.
\end{equation}
The last integral equals a constant independent of $t$, and the $t$-dependent terms in the factor $e^{F_t(0)}$ vanishes if we consider the conjugate kernel $K^{\rm conj}_0$ of $K_0$, see (\ref{ConjKernel}). In fact,
\begin{eqnarray}
e^{F_t(0)}&=& \left(\frac{(d-1)^{d-1}}{d^d}\right)^{-t^{2/3} \kappa^2(r_2-r_1)\frac{2}{d(d-1)}}
\left(\frac{d-1}{d}\right)^{-t^{1/3}\kappa(s_2-s_1)}\frac{d}{d-1} \nonumber \\
&=& \left(\frac{(d-1)^{d-1}}{d^d}\right)^{n_1-n_2}\left(\frac{d-1}{d}\right)^{x_2+d n_2-(x_1+d n_1)}\frac{d}{d-1}.
\end{eqnarray}

Let the $\e$ and the $t$ chosen above be denoted by $\e_0$ and $t_0$ respectively. Then the decay is exponentially small in $(s_1+s_2)^{2/3}$ times a constant independent of $L,\e$ and $t$ for $0<\e\leq \e_0$ and $t\geq t_0$. Thus, the whole result can be simply bounded by $e^{-(s_1+s_2)}$ for $L$ large enough. Therefore the bound (\ref{eqLemmaModerate}) holds.
\end{proofOF}

\subsubsection*{Proof of large deviation bound}
\begin{proofOF}{Proposition~\ref{PropLargeDev}}
In this proof the notation $\simeq$ means that the two expressions are equal up to a factor which is not exponentially large in $t$. More precisely, we say that $f\simeq g$ if $\lim_{t\to\infty}\frac{1}{t}\ln(f/g)=0$. We have
\begin{eqnarray}
K^{\rm conj}_0(n_1,x_1;n_2,x_2)&=&\sum_{k=0}^{n_2-1} \Psi_{n_1-n_2+k}^{n_1}(x_1)\left(\frac{d-1}{d}\right)^{x_1+n_1}\left(\frac{d^d}{(d-1)^{d-1}}\right)^{n_1}\nonumber \\
&\times &\Phi_k^{n_2}(x_2)\left(\frac{d}{d-1}\right)^{x_2+n_2} \left(\frac{(d-1)^{d-1}}{d^d}\right)^{n_2}.
\end{eqnarray}
Let us denote $\alpha\equiv\alpha(k)=k/t$. Then we have
\begin{equation}\label{eq5.68}
K^{\rm conj}_0(n_1,x_1;n_2,x_2)\simeq \sum_{k=0}^{n_2-1} I_1(\alpha(k),\tilde s_1) I_2(\alpha(k),\tilde s_2)
\end{equation}
where
\begin{eqnarray}\label{eq4.63}
I_1(\alpha,s)&\simeq& \frac{1}{2\pi \I}\oint_{\Gamma_{-1}}\dx u \exp\big(t f_{\alpha,s}(u)\big),\nonumber \\
I_2(\alpha,s)&\simeq& \frac{1}{2\pi \I}\oint_{\Gamma_{0}}\dx v \exp\big(-t f_{\alpha,s}(v)\big),
\end{eqnarray}
and
\begin{eqnarray}\label{eq4.64}
f_{\alpha,s}(u)&=&\ln(1+pu)+\alpha\ln(-u)-(d-1)\Big(\frac{p}{d-p}-\alpha\Big)\ln(1+u)\nonumber \\
& &+s\big(\ln(1+u)-\ln(1-1/d)\big).
\end{eqnarray}
We just have to find bounds on $I_1$ and $I_2$ such that their product is exponentially small in $t$. Then, since the sum (\ref{eq5.68}) includes $\Or(t)$ products, the result is obtained by determining the $\alpha\in [0,p/(d-p)]$ which minimizes $I_1(\alpha,s)I_2(\alpha,s)$.

The stationary points of $f_{\alpha,s}(u)$, are by the Cauchy-Riemann equations also the critical points of $\Re(f_{\alpha,s}(u))$. Denote $\beta=(d-1)(p/(d-p)-\alpha)$, then we have to consider only $\alpha>0$, resp.\ $s<\beta$, because the limit cases $\alpha=0$, resp.\ $s=\beta$, correspond to a vanishing pole at $u=0$, resp.\ $u=-1$. Thus when $\alpha=0$, $I_2(\alpha,s)\simeq 0$, and when $s\geq \beta$, $I_1(\alpha,s)\simeq 0$. This is what happens in case (2), because if $\tilde s_i>c=(d-1)p/(d-p)$, then $\beta<s$ for all $\alpha\geq 0$.

First we consider $s=0$. Then
\begin{equation}
\Dt{f_{\alpha,0}(u)}{u}\equiv \frac{p}{1+pu}+\frac{\alpha}{u}-\frac{\beta}{1+u}=0
\end{equation}
has two solutions in $(-1,0)$,
\begin{eqnarray}\label{eq4.67}
u_{-,0}&=&\min\Big\{-\frac{1}{d},-\frac{\alpha(d-p)}{p(\alpha(d-p)+(1-p))}\Big\}\nonumber \\
&\leq& \max\Big\{-\frac{1}{d},-\frac{\alpha(d-p)}{p(\alpha(d-p)+(1-p))}\Big\}=u_{+,0}
\end{eqnarray}
with equality for $\alpha=(1-p)p/(d-p)^2\in (0,p/(d-p))$ for any $p\in (0,1)$ and $d\geq 2$.
In Lemma~\ref{Lemma4} we prove that for $s\in (0,\beta)$, $u_{-,s}\in (-1,-1/d)$ is strictly decreasing in $s$, while $u_{+,s}\in (-1/d,0)$ and is strictly increasing in $s$. $u_{+,s}$ is the left-most maximum of $f_{\alpha,s}(u)$ for $u\in (-1,0)$. Thus, for $s>s_0$,
\begin{equation}\label{eq5.74}
f_{\alpha,s}(u_{+,s})>f_{\alpha,s}(u_{+,s_0}) >f_{\alpha,s_0}(u_{+,s_0})
\end{equation}
because $f_{\alpha,s}(u_{+,s_0})=f_{\alpha,s_0}(u_{+,s_0})+(s-s_0)\big(\ln(1+u_{+,s_0})-\ln(1-1/d)\big)$.
Similarly we obtain, for $s<s_0$, $f_{\alpha,s_0}(u_{-,s_0})<f_{\alpha,s}(u_{-,s}).$

In Lemma~\ref{Lemma5} we prove that $I_1(\alpha,s)\simeq e^{t f_{\alpha,s}(u_{-,s})}$ (resp.\ $I_2(\alpha,s)\simeq e^{-t f_{\alpha,s}(u_{+,s})}$), from which
\begin{equation}
K^{\rm conj}_0(n_1,x_1;n_2,x_2) \simeq \exp\big(t \max_{0\leq \alpha\leq p/(d-p)} (f_{\alpha,\tilde s_1}(v_{-,\tilde s_1})-t f_{\alpha,\tilde s_2}(v_{+,\tilde s_2}))\big).
\end{equation}
Then using (\ref{eq5.74}) we get
\begin{eqnarray}\label{4.68}
K^{\rm conj}_0(n_1,x_1;n_2,x_2)& \lesssim & \exp\Big(t \max_{0\leq \alpha\leq p/(d-p)} (f_{\alpha,\e}(v_{-,\e})-f_{\alpha,\e}(v_{+,\e}))\Big)\nonumber \\
&\lesssim & \exp\Big(-\frac{2}{3}(2\e/\kappa)^{3/2}t+t\Or(\e^2)\Big).
\end{eqnarray}
For $\e\to 0$ the maximum is obtained at $\alpha = (1-p)p/(d-p)^2$. For $\e>0$ small it turns out that the optimization is also obtained for the same $\alpha$. This can be seen by looking at $\alpha+x\sqrt{\e}$ and see that $x=0$ gives the maximum.

Finally, if only one between $\tilde s_1$ and $\tilde s_2$ is in $(\e,c)$, the other being smaller than $\e$, we have to replace $v_{-,\e}$ by $v_{-,0}$ or $v_{+,\e}$ by $v_{+,\e}$. In this case we then obtain the bound \mbox{$\exp\big(-\frac{2}{3}(\e/\kappa)^{3/2}t+t\Or(\e^2)\big)$} instead of the one in (\ref{4.68}). For $0<\e\ll 1$ and $t\gg 1$, we then have the bound
\begin{equation}\label{eq5.77}
\exp\big(-\tfrac{1}{3}(\e/\kappa)^{3/2}t\big).
\end{equation}
Now, since $\e \leq \tilde s_1+\tilde s_2 \leq 2c$, then $t\geq (s_1+s_2)^{3/2} (\kappa/2c)^{3/2}$ and $\sqrt{s_1+s_2}\geq t^{1/3} \sqrt{\e/\kappa}$. Thus we get
\begin{equation}
(\ref{eq5.77}) \leq e^{-(s_1+s_2)^{3/2}(\e/2c)^{3/2}/3} \leq e^{-(s_1+s_2) \e^2 t^{1/3} /(3 c^{3/2} \kappa^{1/2})} \leq e^{-(s_1+s_2)}
\end{equation}
for $t$ large enough.
\end{proofOF}

A remark to point (2) of Proposition~\ref{PropLargeDev}. The kernel is identically equal to zero as soon as the position we look is smaller than the initial position of the particle, thus the maximal value that the $s_i$'s can take without the kernel being identically equal to zero is of order $t$.

\begin{lem}\label{Lemma4}
Let $\alpha>0$, $0<s<\beta$, and consider the function $f_{\alpha,s}$ defined in (\ref{eq4.64}). $F_{\alpha,s}$ has two stationary points at $-1<u_{-,s}<-1/d<u_{+,s}<0$. Moreover, $u_{-,s}$ is strictly decreasing in $s$ and $u_{+,s}$ is strictly increasing in $s$.
\end{lem}
\begin{proof}
In the limit case $s=0$ we have two solutions in $(-1,0)$ but not strictly away from $-1/d$, see (\ref{eq4.67}). First consider $u_{+,s}$. Let us take $s>s_0\geq 0$. By definition $f'_{\alpha,s_0}(u_{+,s_0})=0$ and $\lim_{u\to 0^-}f'_{\alpha,s_0}(u)=-\infty$, therefore $f'_{\alpha,s_0}(u)<0$ for $u\in (u_{+,s_0},0)$. On the other hand,
\begin{equation}
f'_{\alpha,s}(u_{+,s_0})=f'_{\alpha,s_0}(u_{+,s_0})+\frac{s-s_0}{1+u_{+,s_0}}>0,\quad s<s_0.
\end{equation}
Therefore $u_{+,s}\in (u_{+,s_0},0)$, thus $u_{+,s}$ is strictly increasing in $s$. The result for $u_{-,s}$ follows in a similar way. The shape of $\Re(f_{\alpha,s})$ for $\alpha>0,0<s<\beta$ is shown in Figure~\ref{FigLevelLines}.
\begin{figure}[t!]
\begin{center}
\includegraphics[bb=0 300 460 570,clip,height=4cm]{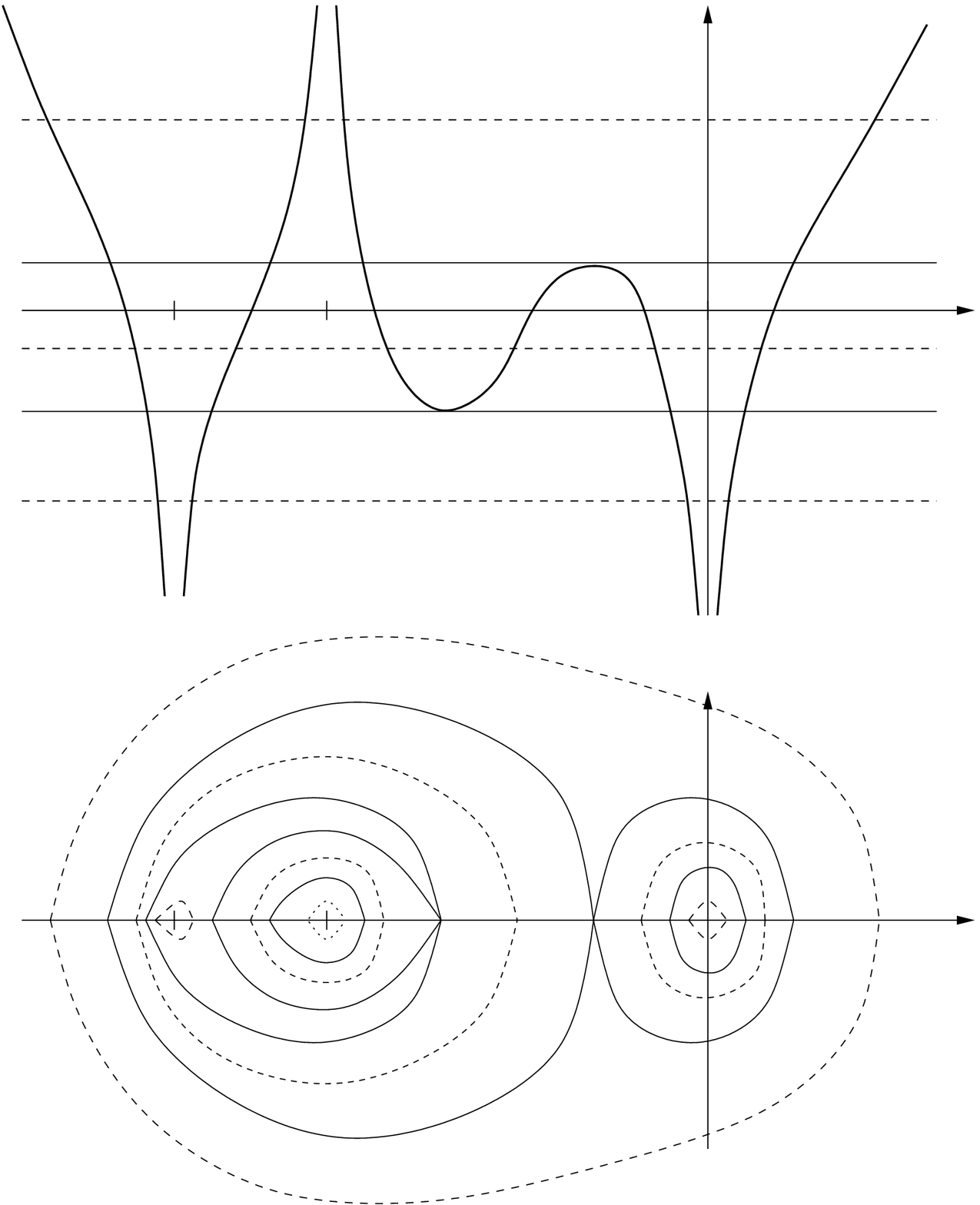}
\caption{Shape of $\Re(f_{\alpha,s})$ along the real axis. The shown divergences are at $-1/p$, $-1$ and $0$.}\label{FigLevelLines}
\end{center}
\end{figure}

\end{proof}

\begin{lem}\label{Lemma5}
Let $\alpha>0$ and $\tilde s_i<\beta$. Then the following asymptotic exponential behaviors hold
\begin{equation}
I_1(\alpha,\tilde s_1)\simeq \exp\big(t f_{\alpha,\tilde s_1}(u_{-,s})\big),
\quad I_1(\alpha,\tilde s_1)\simeq \exp\big(-t f_{\alpha,\tilde s_2}(u_{+,s})\big).
\end{equation}
\end{lem}
\begin{proof}
We will show that $\Gamma_{-1}=\{u=-1+(u_{-,s}+1)e^{i\theta},\theta\in [-\pi,\pi)\}$ and
$\Gamma_{0}=\{v=-u_{+,s}e^{i\theta},\theta\in [-\pi,\pi)\}$ are steep descent paths\footnote{For an integral $I=\int_\gamma \dx z e^{t f(z)}$, we say that $\gamma$ is a steep descent path if (1) $\Re(f(z))$ is maximal at some $z_0\in\gamma$: $\Re(f(z))< \Re(f(z_0))$ for $z\in\gamma\setminus\{z_0\}$ and (2) $\Re(f(z))$ is monotone along $\gamma\setminus\{z_0\}$ except, if $\gamma$ is closed, at a single point where $\Re(f)$ is minimal.} of the integral defining $I_1$ and $I_2$ with maximal value at $\theta=0$. The Lemma follows then from the fact that the paths $\Gamma_{-1}$ and $\Gamma_0$, which pass through $u_{-,s}$ and $u_{+,s}$, are of length of order $1$ in $t$.

Let us start with $I_1(\alpha,s)$. Denote $g(\theta)=\Re(f_{\alpha,s}(u))$ for $u\in \Gamma_{-1}$, and $r=u_{-,s}+1$. Explicitly,
\begin{eqnarray}
g(\theta)&=&\frac12 \ln\big((1-p)^2+p^2r^2+2p(1-p)r\cos(\theta)\big)\nonumber \\
& &+\frac{\alpha}{2}\ln\big(1+r^2-2r\cos(\theta)\big)+\cte_{11}
\end{eqnarray}
for some $\cte_{11}$ independent of $\theta$. Then
\begin{equation}
\Dt{g(\theta)}{\theta} = r\sin(\theta)\Big(\frac{\alpha}{1+r^2-2r\cos(\theta)}-\frac{1}{\frac{1-p}{p}+\frac{p}{1-p}r^2+2r\cos(\theta)}\Big).
\end{equation}
$\theta=0$ corresponds to $u=u_{-,s}$, which is a simple zero of $\Dt{f(u)}{u}$, thus a saddle point. $u=u_{-,s}$ is a local minimum of $\Re(f(u))$ along the real direction, thus it is a local maximum along the path $\Gamma_{-1}$ at $\theta=0$ and for $|\theta|\ll 1$, $\Re(f(u))\simeq -c \theta^2$ for some $c>0$. This implies that the term inside the brackets is strictly negative at $\theta=0$.

To see that $\Gamma_{-1}$ is a steep descent path, we have to check that the term inside the brackets is negative, as already shown for $\theta=0$. Thus we have to show that
\begin{equation}
\frac{1-p}{p}+\frac{p}{1-p}r^2+2r\cos(\theta) < \alpha^{-1}\big(1+r^2-2r\cos(\theta)\big).
\end{equation}
This rewrites as
\begin{equation}
2r\cos(\theta)(1+\alpha^{-1}) < \frac{1}{\alpha}-\frac{1-p}{p}+r^2 \left(\frac{1}{\alpha}-\frac{p}{1-p}\right).
\end{equation}
This inequality holds for all $\theta$ because it holds for $\theta=0$ where the $\cos(\theta)$ is maximal.

Now, consider $I_2(\alpha,s)$. Let $\tilde g(\theta)=\Re(-f_{\alpha,s}(v))$ for $v\in\Gamma_0$ and denote $r=-u_{+,s}$. Then
\begin{equation}
\tilde g(\theta)=-\frac12 \ln(1+p^2r^2-2pr\cos(\theta))+\frac{\beta-s}{2}\ln(1+r^2-2r\cos(\theta))+\cte_{12}
\end{equation}
for some $\cte_{12}$ independent of $\theta$.
Then
\begin{equation}
\Dt{\tilde g(\theta)}{\theta}= r\sin(\theta)\left(\frac{\beta-s}{1+r^2-2r\cos(\theta)}-\frac{1}{p^{-1}+pr^2-2r\cos(\theta)}\right).
\end{equation}
If the term in the parenthesis is negative for all $\theta$, then $\Gamma_0$ is a steep descent path with maximum at $u_{+,s}$. This condition writes
\begin{equation}
1-\frac{\beta-s}{p}+r^2(1-p(\beta-s))>2r\cos(\theta)(1-\beta+s).
\end{equation}
Since $1-\beta+s\geq 1-\beta=1-(d-1)p/(d-p)+(d-1)\alpha \geq 1-(d-1)p/(d-p)>0$ for $p\in (0,1)$, it follows that the inequality holds for all $\theta$ if it holds at $\theta=0$. But there the inequality holds since $u_{+,s}$ is a local quadratic minimum along the real direction of $\tilde g(u)$.
\end{proof}

\subsubsection*{Proof of the bound of the binomial term}
\begin{proofOF}{Proposition~\ref{PropBound}}
We divide the analysis into two cases.\\
1) $|s_1-s_2|\leq t^{1/6}$. In this case the results follows from the estimate (\ref{eq5.15}) with $\Or(\e)\leq \Or(t^{-1/6})$. This, together with the inequality
\begin{equation}\label{eq5.ineq}
e^{-x^2/(4r)}\leq e^{4r} e^{-|x|},
\end{equation}
leads to the desired bound.\\
2) $|s_1-s_2|\geq t^{1/6}$. Denote $a$ as in (\ref{eq5.14}) and $x_1-x_2-1=\alpha a$. $a\simeq t^{2/3}\gg 1$ for large $t$. Then $|s_1-s_2|\geq t^{1/6}$ corresponds to $|\alpha-d|\geq c_1 t^{-1/6}$ with $c_1=d(d-1)/(2\kappa(r_2-r_1))$. The $\alpha$-dependent term in the l.h.s.~of (\ref{eq5.66}) is
\begin{equation}\label{eq5.67}
\left(\frac{d}{d-1}\right)^{-\alpha a} \binom{\alpha a}{a}\simeq e^{a g(\alpha)},
\end{equation}
with
\begin{equation}
g(\alpha)=\left(\alpha\ln(\alpha)-(\alpha-1)\ln(\alpha-1)-\alpha\ln(d/(d-1))\right).
\end{equation}
In (\ref{eq5.67}) the correction to $e^{a g(\alpha)}$ behaves like $1/\sqrt{a}$, thus a prefactor $1/\sqrt{r_2-r_1}$ will always be in $\cte_1$. We have
\begin{equation}
\Dt{g(\alpha)}{\alpha}=\ln(\alpha/(\alpha-1))-\ln(d/(d-1)),\quad \Dtn{g(\alpha)}{\alpha}{2}=-\frac{1}{\alpha(\alpha-1)}.
\end{equation}
Thus $\Dt{g(\alpha)}{\alpha}=0$ only at $\alpha=d$ and $\Dtn{g(\alpha)}{\alpha}{2}$ is increasing in $\alpha$. Thus
for $\alpha\in [1,d]$,
\begin{equation}
g(\alpha)\leq g(d)+f_1(\alpha),\quad f_1(\alpha)=-\frac{1}{2d(d-1)}(\alpha-d)^2.
\end{equation}
Moreover, for $\alpha\in[1,\alpha_c]$,
\begin{equation}
g(\alpha)\leq g(d)+f_2(\alpha),\quad f_2(\alpha)=-\frac{1}{4d(d-1)}(\alpha-d)^2
\end{equation}
where $\alpha_c$ is defined by $f_2'(\alpha_c)=\lim_{\alpha\to\infty}g'(\alpha)$: $\alpha_c=d+2d(d-1)\ln(d/(d-1))$. For $\alpha\geq \alpha_c$, we set $f_3(\alpha)=(\alpha-d) f_2(\alpha_c)/(\alpha_c-d)$. Then $g(\alpha)\leq g(d)+f_3(\alpha)$. With these bounds on $g(\alpha)$ we then get:\\
a) $1\leq \alpha\leq d$: $e^{a g(\alpha)}\leq e^{a g(d)} e^{-(s_2-s_1)^2/(4(r_2-r_1))}$,\\
b) $d\leq \alpha\leq \alpha_c$: $e^{a g(\alpha)}\leq e^{a g(d)} e^{-(s_2-s_1)^2/(8(r_2-r_1))}$,\\
c) $\alpha\geq \alpha_c$: $e^{a g(\alpha)}\leq e^{a g(d)} e^{-|s_2-s_1|t^{1/3} \kappa d_2(\alpha_c)/(2d(d-1)\ln(d/(d-1)))}$.\\
Finally, by (\ref{eq5.ineq}) the desired bound holds for $t$ large enough.
\end{proofOF}

\appendix
\section{Kernel $K_{\rm F_1}$ and trace-class}\label{AppTraceClass}
\subsubsection*{Fredholm's series point of view}
One way of looking at (\ref{eq2.4}) is by simply considering the Fredholm series
\begin{eqnarray*}
& & \det(\Id-\chi_s K_{\rm F_1}\chi_s)_{L^2(\{u_1,\ldots,u_m\}\times\R)}\\
&=& \sum_{n\geq 0}\frac{(-1)^n}{n!} \sum_{i_1,\ldots,i_n=1}^m \int_{s_{i_1}}^\infty\dx x_1 \cdots \int_{s_{i_n}}^\infty \dx x_n \det(K(u_{i_k},x_k;u_{i_l},x_l))_{1\leq k,l\leq n}.
\end{eqnarray*}
Then it is not difficult to see that the function is absolutely sumable/integrable, so that the series is well defined. The required bound is easily obtained if we conjugate the kernel with
\begin{equation}
\rho(u_k,x)=(1+x^2)^{2k}
\end{equation}
as follows. We use
\begin{equation}
\det\left(K(u_{i_k},x_k;u_{i_l},x_l)\right)_{1\leq k,l\leq n}=\det\left(K(u_{i_k},x_k;u_{i_l},x_l) \frac{\rho(u_{i_l},x_l)}{\rho(u_{i_k},x_k)}\right)_{1\leq k,l\leq n}
\end{equation}
and the Hadamard bound, that is, the absolute value of a determinant of a $n\times n$ matrix with entries of absolute value not exceeding $1$ is bounded by $n^{n/2}$. The details are like in the proof of Theorem~\ref{ThmConvToSasProc}.

\subsubsection*{Operator point of view}
The second point of view is to consider (\ref{eq2.4}) as a Fredholm determinant of an operator. The Fredholm determinant is well defined for trace-class operators. What we then mean by (\ref{eq2.4}) is that there exists a \emph{conjugate} operator (i.e., leading to the same determinantal measure) which is trace-class on $\Hilb=L^2(\{u_1,\ldots,u_m\}\times\R)$.

Consider the conjugate operator $\widetilde K_1$ given by the kernel
\begin{equation}
\widetilde K_1(u_1,s_1;u_2,s_2)=K_1(u_1,s_1;u_2,s_2) \frac{\rho(u_2,s_2)}{\rho(u_1,s_1)}.
\end{equation}
The choice of $\rho(u_k,x)$ is not at all unique. We use the following one:
\begin{equation}
\rho(u_k,x)=(1+x^2)^{2k}.
\end{equation}

\begin{prop}\label{PropTraceClass}
The operator $\widetilde K_1$ as defined above is trace-class on \mbox{$\Hilb=L^2(\{u_1,\ldots,u_m\}\times\R)$}.
\end{prop}
\begin{proof}
We can write $\widetilde K_1=\sum_{k,l=1}^m P_k \widetilde K_1 P_l$ where $P_k$ is the projection from $\Hilb$ into its subspace $\Hilb_k=\{f\in\Hilb| f(u_i,x)=0,i\neq k\}$. Then
\begin{equation}
\|\widetilde K_1\|_{1,\Hilb}\leq \sum_{k,l=1}^m \|P_k \widetilde K_1 P_l\|_{1,\Hilb}.
\end{equation}
Now, $\|P_k \widetilde K_1 P_l\|_{1,\Hilb}=\|\widetilde K_1^{k,l}\|_1$ where
$\widetilde K_1^{k,l}$ is the operator on $L^2(\R)$ with kernel $\widetilde K_1(u_k,\cdot;u_l,\cdot)$. Thus to prove that $\widetilde K_1$ is trace-class on $\Hilb$, it is enough to prove that, for $k,l=1,\ldots,m$, $\widetilde K_1^{k,l}$ is trace-class as operator on $L^2(\R)$. This is proven in Lemma~\ref{LemTrace1} and Lemma~\ref{LemTrace2}.
\end{proof}

\begin{lem}\label{LemTrace1}
The operator with kernel
\begin{equation}
L_{k,l}(x,y)=\big(e^{(u_k-u_l)\Delta}\big)(x,y)\frac{\rho(u_k,x)}{\rho(u_l,y)}\Id_{[x\geq s_k]}\Id_{[y\geq s_l]}\Id_{[u_k>u_l]}
\end{equation}
is trace-class on $L^2(\R)$.
\end{lem}
\begin{proof}
For $u_k\leq u_l$ nothing has to be proven since $L_{k,l}=0$. Consider $u_k>u_l$ (i.e., $k>l$) and define the two operators
\begin{equation}
A(x,z)=\big(e^{\tfrac12(u_k-u_l)\Delta}\big)(x,z)\frac{(1+z^2)^{k+l}}{(1+x^2)^{2k}}
\end{equation}
and
\begin{equation}
B(z,y)=\big(e^{\tfrac12(u_k-u_l)\Delta}\big)(z,y)\frac{(1+y^2)^{2l}}{(1+z^2)^{k+l}}.
\end{equation}
Then
\begin{equation}
L_{k,l}=P_{s_k} A B P_{s_l}
\end{equation}
with $P_a(x)=\Id_{[x\geq a]}$. The $P$'s are projectors, thus $\|P_{s_\cdot}\|_\infty=1$. From this it follows
\begin{equation}
\|L_{k,l}\|_1\leq \|A\|_2 \|B\|_2.
\end{equation}

It is simple to prove that $\|A\|_2< \infty$ and $\|B\|_2<\infty$, since the $2$-norm is easily bounded using its integral kernel.
\begin{equation}
\|A\|_2^2 \leq \int_{\R^2}\dx x\dx z \frac{1}{2\pi(u_k-u_l)} e^{-2(x-z)^2/(u_k-u_l)}\frac{(1+z^2)^{2(k+l)}}{(1+x^2)^{2(k+l)}} \frac{1}{(1+x^2)^{2(k-l)}}
\end{equation}
By changing the variable $z=y+x$, it is easy to see that
\begin{equation}
\int_\R\dx z \frac{1}{2\pi(u_k-u_l)} e^{-2(x-z)^2/(u_k-u_l)}\frac{(1+z^2)^{2(k+l)}}{(1+x^2)^{2(k+l)}} \leq \cte_{13}
\end{equation}
Therefore
\begin{equation}
\|A\|_2^2 \leq \cte_{13} \int_{\R}\dx x \frac{1}{(1+x^2)^{2(k-l)}} \leq \frac{\pi}{2} \cte_{13}
\end{equation}
since $k-l\geq 1$. In the same way we can check that $\|B\|_2<\infty$, thus proving the result of the Lemma.
\end{proof}

\begin{lem}\label{LemTrace2}
The operator with kernel
\begin{equation}
M_{k,l}(x,y)=\big(e^{-u_l\Delta} B_0 e^{u_k\Delta}\big)(x,y)\frac{\rho(u_k,x)}{\rho(u_l,y)}\Id_{[x\geq s_k]}\Id_{[y\geq s_l]}
\end{equation}
is trace-class on $L^2(\R)$.
\end{lem}
\begin{proof}
First of all, let $P_a$ be the projector on $[a,\infty)$, and set
\begin{equation}
Q_{k,l}(x,y)=\big(e^{-u_l\Delta} B_0 e^{u_k\Delta}\big)(x,y)\frac{\rho(u_k,x)}{\rho(u_l,y)}.
\end{equation}
Then, for $-L\leq \min\{u_k,u_l\}$, it holds that
\begin{equation}
M_{k,l}=P_{u_k} Q_{k,l} P_{u_l} = P_{u_k} P_{-L} Q_{k,l} P_{-L} P_{u_l}
\end{equation}
and, using $\|P_a\|_\infty=1$, we get
\begin{equation}
\|M_{k,l}\|_1\leq \|P_{-L} Q_{k,l} P_{-L}\|_1.
\end{equation}
Instead of using the projectors, we can think of $Q_{k,l}$ as an operator on \mbox{$\Hilb_L=L^2([-L,\infty),\dx x)$}. The Airy function can be expressed as
\begin{equation}
\Ai(x)=\frac{1}{2\pi}\int_{\R}\dx \xi e^{-\xi^2\sigma} e^{\tfrac13 \sigma^3-x\sigma}e^{i(\tfrac13 \xi^3-\xi\sigma^2+x\xi)}
\end{equation}
for any $\sigma>0$. Let us set
\begin{equation}
\alpha(\xi)=\tfrac13 \xi^3-\xi\sigma^2+(u_k-u_l)^2\xi,\quad \beta=\tfrac13\sigma^3-(u_k-u_l)^2\sigma+\tfrac23 (u_k-u_l)^3
\end{equation}
and define the measures
$\dx\mu(x)=(1+x^2)^{-2l}\dx x$, $\dx \nu(x)=(1+x^2)^{2k}\dx x$. With these notations we have, for $f\in\Hilb_L$,
\begin{equation}\label{eq318P}
|\langle f,Q_{k,l} f\rangle_{\Hilb_L}| = \left|\int_{-L}^\infty\dx \mu(x) \int_{-L}^\infty\dx \nu(y) \overline{f(x)} Q_{k,l}(x,y) f(y)\right|
\end{equation}
and
\begin{equation}\label{eqQkl}
Q_{k,l}(x,y)=\frac{e^{\beta}}{2\pi}\int_{\R}\dx \xi e^{-\xi^2\sigma} e^{i\alpha(\xi)} e^{(x+y)(i\sigma-\tau)},
\end{equation}
where we set $\tau=-(u_k-u_l-\sigma)$. By choosing $\sigma>u_k-u_l$ we have $\tau>0$ and we can exchange the integrals. Let us set
\begin{equation}
g_1(\xi,\sigma)=\int_{-L}^\infty\dx \mu(x) \overline{f(x)} e^{ix\xi} e^{-\tau x},\quad
g_2(\xi,\sigma)=\int_{-L}^\infty\dx \nu(x) f(x) e^{i x\xi} e^{-\tau x}.
\end{equation}
Then
\begin{eqnarray}\label{eq321P}
|\langle f,Q_{k,l} f\rangle_{\Hilb_L}| & \leq & \frac{e^{\beta}}{2\pi}\int_{\R}\dx \xi e^{-\xi^2\sigma} \left|g_1(\xi,\sigma)\right| \, \left| g_2(\xi,\sigma)\right| \nonumber \\
& \leq & \frac{e^\beta}{4\pi}\int_{\R}\dx \xi e^{-\xi^2\sigma} (|g_1(\xi,\sigma)|^2+|g_2(\xi,\sigma)|^2)
\end{eqnarray}
where we used $a^2+b^2\geq 2ab$. The first term can be evaluated as follows.
\begin{eqnarray}
& & \int_{\R}\dx \xi e^{-\xi^2\sigma} |g_1(\xi,\sigma)|^2\\
&=&\int_{\R}\dx \xi e^{-\xi^2\sigma}\left(\int_{-L}^\infty\dx \mu(x) f(x) e^{-ix\xi} e^{-\tau x}\right)
\left(\int_{-L}^\infty \dx\mu(y) \overline{f(y)} e^{iy\xi} e^{-\tau y}\right)\nonumber \\
&=&\int_{-L}^\infty\dx \mu(x)\int_{-L}^\infty\dx \mu(y)\overline{f(y)} f(x) e^{-\tau (x+y)} \int_{\R}\dx \xi e^{-\xi^2\sigma} e^{i(x-y)\xi}\nonumber \\
&=&\sqrt{\frac{\pi}{\sigma}}\int_{-L}^\infty\dx \mu(x)\int_{-L}^\infty\dx \mu(y)\overline{f(y)} f(x) e^{-\tau (x+y)} e^{-(x-y)^2/4\sigma}.\nonumber
\end{eqnarray}
The second term is computed in essentially the same way. We obtain
\begin{equation}\label{eqG1}
0\leq |\langle f,Q_{k,l} f\rangle_{\Hilb}| \leq \frac{e^{\beta}}{4\sqrt{\pi\sigma}}\langle f,G f\rangle_{\Hilb}
\end{equation}
with $G$ the operator with kernel
\begin{equation}\label{eqG2}
G(x,y)=e^{-\tau(x+y)}e^{-(x-y)^2/(4\sigma)} \left((1+x^2)^{2k}(1+y^2)^{2k}+(1+x^2)^{-2l}(1+y^2)^{-2l}\right).
\end{equation}

$Q_{k,l}$ is Hilbert-Schmidt on $\Hilb_L$ (use (\ref{eqQkl})). Thus $Q_{k,l}$ is a compact operator. For any orthonormal basis $\{\phi_n\}_{n\geq 1}$ we have
\begin{equation}
\sum_{n\geq 1}|\langle \phi_n, Q_{k,l} \phi_n\rangle_{\Hilb_L}|
\leq \frac{e^{\beta}}{4\sqrt{\pi\sigma}} \sum_{n\geq 1}\langle \phi_n, G \phi_n\rangle_{\Hilb_L}.
\end{equation}
By (\ref{eqG1}) and (\ref{eqG2}) $G$ is a positive and continuous operator, thus we have
\begin{equation}
\sum_{n\geq 1}\langle \phi_n, G \phi_n\rangle_{\Hilb_L}= \int_{-L}^\infty\dx x G(x,x)=\int_{-L}^\infty \dx x e^{-2\tau x} ((1+x^2)^{4l}+(1+x^2)^{-4k})<\infty.
\end{equation}
This holds \emph{for any} orthonormal basis, thus the Lemma is proved.
\end{proof}

\section{Krawtchouk orthogonal polynomials}\label{SectAppendix}
Let $K_n(x,p,t)= \phantom{}_2F_1(-n,-x;-t;1/p)$ be the Krawtchouk polynomials, see e.g.~\cite{KS96}.
To obtain an integral representation we use their generating function
\begin{equation}
\sum_{n=0}^T \binom{T}{n} K_n(z,p,T) \zeta^n = \left(1-\frac{1-p}{p}\zeta\right)^z (1+\zeta)^{T-z}.
\end{equation}
From this it follows that
\begin{equation}
\binom{T}{n} K_n(z,p,T)=\frac{1}{2\pi \I} \oint_{\Gamma_0}d\zeta \frac{\left(1-\frac{1-p}{p}\zeta\right)^z (1+\zeta)^{T-z}}{\zeta^{n+1}}
\end{equation}
where $\Gamma_0$ is an anticlockwise oriented simple loop with $0$ as the only pole inside $\Gamma_0$.

In the case of particles starting from $y_j=-d(j-1)$ we obtain, for $k=0,\ldots,N-1$,
\begin{equation}\label{eqA3}
\Psi_k^{N}(z)=(1-p)^t \binom{t+k}{z-(d-1)k}\left(\frac{p}{1-p}\right)^{z-(d-1)k} K_k(z-(d-1)k,p,t+k).
\end{equation}
We use the two identities
\begin{eqnarray}
& &(x+1) K_n(x,p,N)\\
& &=(N+1) p \big(K_n(x+1,p,N+1)-K_{n+1}(x+1,p,N+1)\big)\nonumber
\end{eqnarray}
and
\begin{eqnarray}
& &(N+1-x) K_n(x,p,N)\\
& &=(N+1) (1-p) \big(K_n(x,p,N+1)+\frac{p}{1-p} K_{n+1}(x,p,N+1)\big)\nonumber
\end{eqnarray}
recursively and eventually obtain, setting $T=t+d(N-1)$,
\begin{equation}
\Psi_k^{N}(z) = \omega_{T}(z) \sum_{l=0}^{d(N-1)} S_{k,l} K_l(z,p,T)
\end{equation}
with \begin{equation}
\omega_{T}(z)=(1-p)^{T}\left(\frac{p}{1-p}\right)^{z} \binom{T}{z}
\end{equation}
the standard weight for the Krawtchouk polynomials $K_j(z,p,T)$, \mbox{$j=0,\ldots,T$}, and the (not square) matrix $S$ with entries
\begin{equation}
S_{i,j}=\left(\frac{p}{1-p}\right)^{j} \frac{1}{p^i} \sum_{\lambda\geq 0} \binom{(d-1) i}{\lambda} \binom{d(N-1-i)}{j-i-\lambda} \left(\frac{-p}{1-p}\right)^{-\lambda}.
\end{equation}

To obtain a family of polynomials $\{\Phi_k^{N}(z),k=0,\ldots,N-1\}$, with $\Phi_k^{N}$ of degree $k$, and which
satisfy $\sum_{z\geq 0}\Phi_k^{N}(z)\Psi_j^{N}(z)=\delta_{k,j}$, we need to be able to invert the matrix $S$ restricted to the entries $0\leq i,j\leq N-1$. The orthogonality relation of the Krawtchouk polynomials is, for $0\leq n,m\leq T$,
\begin{equation}
\sum_{z=0}^T \omega_T(z) K_m(z,p,T) K_n(z,p,T)=\left(\binom{T}{n}\left(\frac{p}{1-p}\right)^{n}\right)^{-1} \delta_{m,n}.
\end{equation}
Therefore
\begin{equation}
\Phi_k^{N}(z) = \sum_{l=0}^{N-1}K_l(z,p,T) \binom{T}{l} \left(\frac{p}{1-p}\right)^l S^{-1}_{l,k}.
\end{equation}

In the inverse of the matrix $S=[S_{i,j}]_{0\leq i,j,\leq N-1}$, the terms in front of the sum are easily accounted for. In fact, if a matrix $S$ has entries $S_{i,j}=f(i)\tilde S_{i,j} g(j)$, then its inverse has entries $S^{-1}_{i,j}=g^{-1}(i) \tilde S^{-1}_{i,j} f^{-1}(i)$. In our case, $f(i)= p^{-i}$ and $g(j)=\big(\frac{p}{1-p}\big)^{j}$, and $S_{i,j}=\sum_{\lambda\geq 0} \binom{(d-1) i}{\lambda} \binom{d(N-1-i)}{j-i-\lambda}\xi^\lambda$. with $\xi=-\frac{1-p}{p}$.

\subsection*{Case $d=2$}
In this case, set $T=t+2(N-1)$ and $\omega_{T}(z) =(1-p)^{T}\left(\frac{p}{1-p}\right)^{z} \binom{T}{z}$. Then
\begin{equation}
S_{i,j} = \left(\frac{p}{1-p}\right)^{j} \frac{1}{p^i} \sum_{\lambda\geq 0}^{(j-i)}\left(\frac{1-p}{-p}\right)^{\lambda} \binom{i}{\lambda} \binom{2(N-1-i)}{j-i-\lambda},
\end{equation}
\begin{equation}
\Psi_k^{N}(z) = \omega_{T}(z) \sum_{l=0}^{2(N-1)} S_{k,l} K_l(z,p,T),
\end{equation}
and
\begin{equation}
\Phi_k^{N}(z) = \sum_{l=0}^{N-1}K_l(z,p,T) \binom{T}{l} \left(\frac{p}{1-p}\right)^l S^{-1}_{l,k}.
\end{equation}
The computation of the inverse of $S$ is quite involved, but at the end of the week one finds, with $0\leq i,j \leq N-1$,
\begin{equation}
S^{-1}_{i,j}=\left(\frac{1-p}{p}\right)^i \! p^j \sum_{\lambda\geq 0}^{(j-i)} \left(\frac{1-p}{-p}\right)^{\lambda} \! (-1)^{j-i} \binom{2N-2-i-j-\lambda}{j-i-\lambda}\binom{j+\lambda}{\lambda}A^{(\lambda)}_{i,j}
\end{equation}
with
\begin{eqnarray*}
A^{(j-i)}_{i,j}&=& \frac{i}{2j-i}, \textrm{ and setting it to be }1\textrm{ for }i=j=0,\\
A^{(\lambda)}_{i,j}&=& 2\frac{j(N-1-i)-\lambda (N-1)}{(j+\lambda)(2N-2-i-j-\lambda)}
\end{eqnarray*}
the last being the expression for $0\leq \lambda \leq j-i-1$.

To obtain the functions $\Phi_k^{N}$ we plug the solution of the inverse of the matrix in the expression above and perform the sums. First one exchanges the path integral and the finite sums. Then the sums can be simplified by extending them to $-\infty$ or $\infty$ depending on the case. This can be done because the extra terms turn out to be analytic functions leading to a zero contribution after integration. In the end we obtain $\Phi_0^{N}=1$ and, for \mbox{$1\leq k\leq N-1$},
\begin{equation}
\Phi_k^{N}(z)=\frac{(-1)^k}{2\pi \I}\oint_{\Gamma_0}\frac{dv}{v} \! \frac{1}{v^k}(1-pv)^{t+2k-z-1} \left(1+(1-p)v\right)^{z-k-1}\left(1+(2-p)v\right).
\end{equation}


\end{document}